\documentclass[usenatbib,usegraphicx]{mn2e}
\usepackage{amsmath}

\usepackage{graphicx}
\usepackage{subcaption}

\usepackage[
dvipdfm,
pdfauthor={Laszlo Szucs},
pdftitle={A benchmark of CO as cloud mass indicator},
pdfstartview=FitH,
linkcolor=blue,
anchorcolor=blue,
citecolor=blue,
filecolor=blue,
menucolor=blue,
urlcolor=blue,
colorlinks=true]{hyperref}

\usepackage{aas_macros}

\title[A benchmark of CO as cloud mass indicator]{How well does CO emission measure the H$_{2}$ mass of MCs?}
\author[Sz\H{u}cs et al.]{ {L\'{a}szl\'{o} Sz\H{u}cs}$^{1, 2}$\thanks{Member of IMPRS for Astronomy \& Cosmic Physics at the
University of Heidelberg.}\thanks{E-mail: laszlo.szucs@mpe.mpg.de},
       {Simon C.~O. Glover}$^{1}$ \& {Ralf S. Klessen}$^{1}$  \\
       $^{1}$ {Zentrum f\"ur Astronomie der Universit\"at Heidelberg, Institut f\"ur Theoretische Astrophysik, Albert-Ueberle-Str.\ 2, 69120 Heidelberg, Germany} \\
       $^{2}$ {Max-Planck-Institut f\"{u}r extraterrestrische Physik, Giessenbachstrasse 1, 85748 Garching, Germany}
}

\newenvironment{itemize*}
  {\begin{itemize}
    \setlength{\itemsep}{0pt}
    \setlength{\parskip}{0pt}}
  {\end{itemize}}

\begin{document}

\maketitle

\begin{abstract}
We present numerical simulations of molecular clouds (MCs) with self-consistent CO gas-phase and isotope chemistry in various environments. The simulations are post-processed with a line radiative transfer code to obtain $^{12}$CO and $^{13}$CO  emission maps for the $J=1\rightarrow0$ rotational transition. The emission maps are analysed with commonly used observational methods, i.e. the $^{13}$CO column density measurement, the virial mass estimate and the so-called $X_{\textrm{CO}}$ (also CO-to-H$_2$) conversion factor, and then the inferred quantities (i.e. mass and column density) are compared to the physical values.
We generally find that most methods examined here recover the CO-emitting H$_{2}$ gas mass of MCs within a factor of two uncertainty if the metallicity is not too low. The exception is the $^{13}$CO column density method. It is affected by chemical and optical depth issues, and it measures both the true H$_{2}$ column density distribution and the molecular mass poorly. The virial mass estimate seems to work the best in the considered metallicity and radiation field strength range, even when the overall virial parameter of the cloud is above the equilibrium value. This is explained by a systematically lower virial parameter (i.e. closer to equilibrium) in the CO-emitting regions; in CO emission, clouds might seem (sub-)virial, even when, in fact, they are expanding or being dispersed. A single CO-to-H$_{2}$ conversion factor appears to be a robust choice over relatively wide ranges of cloud conditions, unless the metallicity is low. The methods which try to take the metallicity dependence of the conversion factor into account tend to systematically overestimate the true cloud masses.
\end{abstract}
\begin{keywords}
   astrochemistry -- hydrodynamics -- radiative transfer -- ISM: abundances -- radio lines: ISM
\end{keywords}

\section{Introduction} \label{sec:intro}

The masses of molecular clouds and their distribution are fundamental parameters for the understanding of the structure of the interstellar medium (ISM) and the star formation process itself. Often, emission from the millimetre-wave rotational transitions of carbon monoxide (CO, hereafter we indicate the isotope mass when referring to a specific isotopologue) and the mass or column density inference methods built upon its measurement provides the most accessible estimate of these fundamental quantities. The most common methods are (1) the local thermodynamic equilibrium (LTE) column density measurement of an optically thin CO isotope \citep[e.g.][]{Goldsmith2008,Pineda2008,Pineda2010}, (2) the virial mass estimate \citep[e.g.][]{Solomon1987,MacLaren1988,Dame2001,Hughes2010} and (3) the direct conversion of the observed $^{12}$CO intensity into H$_2$ mass by applying a (usually fixed) conversion factor \citep[often called $X_{\textrm{CO}}$, see the review of][]{Bolatto2013}. The molecular mass estimates are then used to infer the star formation efficiency \citep[e.g.][]{SolomonSage1988,Kennicutt1998}, the galactic and extragalactic molecular cloud mass functions \citep{Colombo2014,Zaragoza2014} and the ratio of kinetic and gravitational energies \citep[e.g][]{Solomon1987,Colombo2014}, thus inferring the potential collapse or dispersal of individual clouds \citep{Kauffmann2013}. The measured column density distribution may provide an insight into the Mach-number of the turbulent velocity field that shaped it \citep{Padoan1997,Federrath2008,Federrath2010a,Brunt2010,Burkhart2012,Molina2012,Konstandin2015}. Furthermore, CO emission provides the primary measure and the basis of our current understanding of the distribution and kinematics of molecular gas within our Galaxy \citep{HeyerDame2015}.
Any uncertainty in the mass and column density measurements are expected to propagate to these more complex quantities.
Therefore, the reliability of the mass and column density measurement methods, especially any potential systematic effects introduced by the various simplifications and assumptions intrinsic to the different methods, remains a major issue in the field of star formation.

Until recently, the nearby molecular clouds (e.g. Taurus and Perseus), which are well characterised by multiple tracers, provided the best testing grounds for constraining the intrinsic uncertainties in the methods.
For instance, \citet{Pineda2008} used high dynamic range visual extinction maps to determine the H$_{2}$ column density and $^{12}$CO, $^{13}$CO and C$^{18}$O millimetre observations to constrain the CO-to-H$_{2}$ conversion factor and the $^{13}$CO/H$_{2}$ abundance ratio in the Perseus molecular cloud \citep[see also][]{Lee2014}. They show that the $^{12}$CO and $^{13}$CO emission saturates at visual extinctions of 4 mag and 5 mag respectively and that about 60\% of the emission emerges from sub-thermally excited regions. These results clearly challenge the assumptions regarding optical depths and LTE conditions upon which some measurement methods strongly rely.
Such regional studies are, however, restricted to a few nearby molecular clouds and they only probe a narrow range of physical conditions and environments. CO-based techniques, on the other hand, are used in diverse conditions, from low metallicity dwarf galaxies to starburst environments. In addition, the interpretation of observational studies is difficult, since these studies themselves are biased by the effects they ought to measure.

The recent advances in computer power and simulation techniques as well as in the understanding of the physical and chemical processes in the ISM \citep[for an overview, see e.g. the Saas Fee lecture notes by ][]{KlessenGlover2016}, enable increasing realism in numerical simulations. Comparing the true column densities and cloud masses from such simulations to those derived from the simulations by applying the observational methods, might provide the most complete census on the sources of errors affecting the measurements.

There is a large body of work using high-resolution numerical simulations to identify and quantify potential errors and uncertainties in observational assumptions. In one of the first of such studies, \citet{Padoan2000} computed turbulent magneto-hydrodynamical (MHD) simulations with supersonic and super-Alfv\'{e}nic turbulence and post-processed them with non-LTE radiative transfer to obtain CO emission maps. The emission was converted to column density using LTE analysis and then it was compared to the true column density from the simulation. The study finds that none of the assumptions in the LTE method holds over most of the simulated volume, and that the method systematically underestimates the true column densities by a factor of a few. Similar conclusions have been reached by \citet{Shetty2011a,Shetty2011b} when studying the $X_{\textrm{CO}}$ factor, and by \citet{Beaumont2013} in the context of clump analysis. Applications to other types of molecular cloud statistics have been reported for instance by \citet[][]{Bertram2014,Bertram2015a,Bertram2015b}, \citet{Burkhart2013a,Burkhart2013b,Burkhart2013c}, \citet{Chira2014}, \citet{Correia2014}, \citet{Gaches2015}, \citet{Glover2015a,GloverClark2016}, \citet{Walch2015} and \citet{Yeremi2014}.

Here, we present a new investigation of the uncertainties involved in estimating the mass of molecular clouds. Our study is based on turbulent hydrodynamic simulations with self-gravity, an improved approximation of interstellar radiation field (ISRF) attenuation and self-consistent models for chemistry and gas heating/cooling. In addition to the LTE column density estimate technique, we extend our investigation to additional CO-based methods, such as virial mass estimate and the CO-to-H$_{2}$ conversion factor. These methods are calibrated for nearby, ``Solar-like'' or ``Milky Way-like'' clouds, thus they are expected to work reasonably well for such conditions. To test the applicability of the methods for different conditions, we explore a range of molecular cloud parameters (metallicity, cloud mass, virial state) and environments (ISRF strength).

The paper is organised as follows: In section~\ref{sec:sims} and section~\ref{sec:radtran} we describe the basic details of the hydrodynamic simulations and the adopted radiative transfer model, respectively. Section~\ref{sec:massmethods} gives a short overview on the CO-based mass measurements methods and provides details on how the techniques are applied to the synthetic emission maps. The results are presented in section~\ref{sec:result} and in Fig.~\ref{fig:masscomp}. In the discussion (section~\ref{sec:dis}) we pinpoint the reasons for the success or the failure of each of the methods and try to put the results into a broader context. Finally, section~\ref{sec:sum} closes the paper with a summary.

\begin{table*}
\caption{Model parameters and the analysed snapshots}
\label{models}
\begin{center}
\begin{tabular}{ccccccc}
  \hline
  \noalign{\vskip 0.5mm}
  Model & $n_{0}$ [$\textrm{cm}^{-3}$] & Metallicity [$Z_{\odot}$] & ISRF [$G_{0}$] & $\alpha_{\textrm{vir}}$ & Time [Myr] \\
  \noalign{\vskip 0.5mm}
  \hline
  \hline
  \noalign{\vskip 0.5mm}
      \multicolumn{6}{ c }{$M_{\textrm{cloud}} = 10^{4} \, \textrm{M}_{\odot}$} \\
  \noalign{\vskip 0.5mm}
  \hline
      a & 300 & 0.3 & 1 & 1.08 & 2.046   \\
      b & 300 & 0.6 & 1 & 1.10 & 1.930   \\
      c & 300 & 1   & 0.1 & 1.04 & 2.124 \\
      d, d2 & 300 & 1   & 1 & 1.04 & 2.140   \\
      e & 300 & 1   & 10 & 1.07 & 2.022  \\
      f & 1000 & 1  & 1 & 1.51 & 0.973   \\
      g & 300 & 1  & 1 & 2.52 & 2.140   \\
      h & 300 & 1  & 1 & 7.95 & 2.140   \\
  \hline
  \noalign{\vskip 0.5mm}
       \multicolumn{6}{ c }{$M_{\textrm{cloud}} = 10^{5} \, \textrm{M}_{\odot}$} \\
  \noalign{\vskip 0.5mm}
  \hline
      i & 100 & 1    & 1 & 0.90 & 1.297  \\
      j & 100 & 1    & 100 & 0.89 & 1.374  \\
      k & 100 & 0.3  & 1 & 0.81 & 1.445 \\
  \hline
\end{tabular}

\end{center}
\begin{flushleft}
Summary of model parameters. From a) to f) and from i) to k) the analysed snapshots are the last before sink particle formation, while the g) and h) snapshots are taken at the same time as the fiducial case (model d). $Z_{\odot}$ and $G_{0}$ refer to the solar metallicity and the Draine radiation field strength (1.7 in units of the \citet{Habing1968} field), respectively. The virial parameter, $\alpha_{\textrm{vir}}$ is defined as the ratio of 2 times the kinetic energy and the absolute value of the potential energy of the cloud at the given time. Model d2) has the same parameters as the fiducial model, but a different random seed is used to generate the initial turbulent velocity field.
\end{flushleft}
\end{table*}

\section{Simulations} \label{sec:sims}

\begin{figure*}
\begin{center}
\includegraphics[trim=0 170 0 0,scale=0.64]{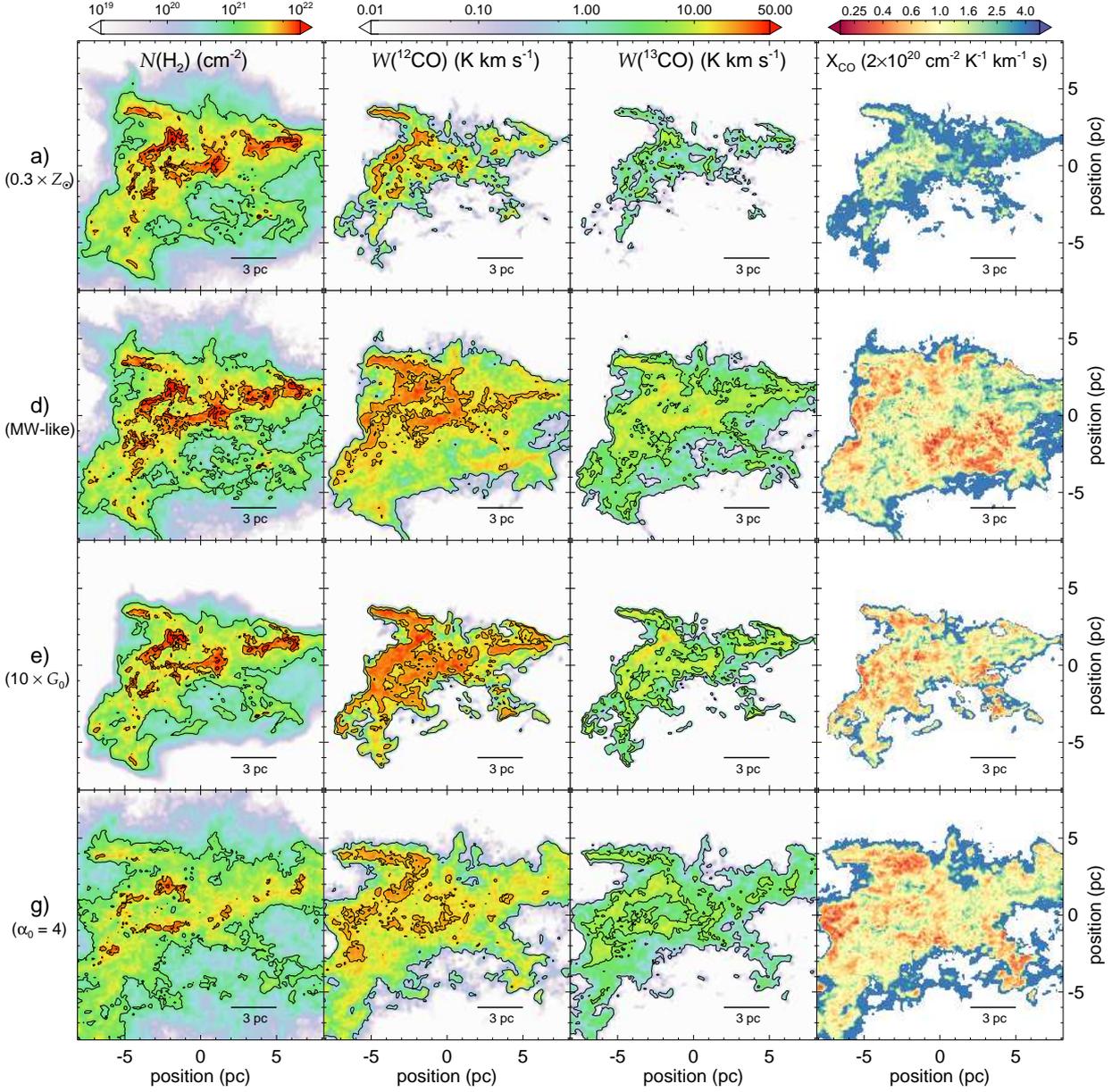}
\end{center}
\caption{H$_{2}$ column density, $^{12}$CO and $^{13}$CO integrated emission maps and the true $X_{\textrm{CO}}$ factor map for selected models (a, d, e and g). The complete set of models is available in the electronic-only material. The contour lines on the H$_{2}$ maps, from outside in, indicate hydrogen molecule column densities of $10^{21}$, $5 \times 10^{21}$ and $10^{22}\,\textrm{cm}^{-2}$. The $^{12}$CO integrated intensity contours indicate the $0.6\,\textrm{K km s}^{-1}$ and $10\, \textrm{K km s}^{-1}$ levels, while the $^{13}$CO contours indicate $W(^{13}\textrm{CO})$ levels of $0.3\,\textrm{K km s}^{-1}$ and $5\,\textrm{K km s}^{-1}$. }
\label{fig:maps}
\end{figure*}

We perform turbulent molecular cloud simulations using the smoothed particle hydrodynamics (SPH) code {\sc GADGET-2}\footnote{\protect\url{http://www.mpa-garching.mpg.de/gadget/}} \citep{Springel2005}. In addition to the base code, which deals with the hydrodynamics of self-gravitating gas, we also include gas-phase chemistry with radiative heating and cooling \citep{GloverClark2012b}, an approximate treatment for the ISRF attenuation \citep{Clark2012a} and sink particle formation \citep{Bate1995,Jappsen2005,Federrath2010b}. For the detailed description of the code, the considered physical and chemical processes and the simulation setup see \citet{Szucs2014}. However, we summarise the most important technical details here as well and underline the differences in initial conditions compared to the above mentioned work. The column density, $^{12}$CO and $^{13}$CO ($J=1\rightarrow0$) integrated emission, and the $X_{\textrm{CO}}$ factor maps from the selected simulations are shown in Fig.~\ref{fig:maps}.

The modelled molecular clouds sample a range of initial number densities, metallicities, incident ISRF strengths, virial parameters and total cloud masses (see Table~\ref{models}). Each simulation starts with a spherical cloud, which has a radius set by the total cloud mass ($\textrm{M}_{\textrm{cloud}}$) and a uniform number density ($n_{0}$). The initial gas and dust temperatures are also uniform at 20 K and 10 K, respectively.
The initial turbulent bulk motions follow a steep, $\textrm{P}(k) \propto k^{-4}$ power spectrum over a factor of 16 in dynamic range in the wavenumber ($k = 2 \pi / l$). The smallest wavenumber (i.e. the largest length, $l$) is equivalent to the respective cloud diameter. The amplitude of the turbulent power spectrum is scaled to give initial virial parameters of $\alpha_{0} = 2, 4, 8$ for the $M_{\textrm{cloud}} = 10^4 M_{\odot}$ and $\alpha_{0} = 1$ for the $M_{\textrm{cloud}} = 10^5 M_{\odot}$ models. The virial parameter is defined as
\begin{equation}
\alpha_{\textrm{vir}} = \frac{2 \times E_{\textrm{kin}}}{|E_{\textrm{pot}}|},
\label{eq:vir}
\end{equation}
where $E_{\textrm{kin}}$ and $E_{\textrm{pot}}$ are the total kinetic and potential energies of the cloud, respectively.
The energy deposited at the top of the hierarchy (small $k$) cascades down quickly to smaller lengths scales (large $k$) and a power spectrum consistent with fully developed turbulence builds up. The turbulent energy is allowed to dissipate through shocks, providing an important mechanism for heating the gas. We do not replenish the dissipated turbulent energy, i.e. turbulent driving is not considered.

The chemical evolution of the turbulent gas is followed within the hydrodynamic simulation using the gas-phase chemical network of \citet[][NL99 hereafter]{NelsonLanger99}, complemented with a simple model for H$_{2}$ formation and destruction \citep[][]{GloverMacLow2007a}. This network is designed to model the carbon chemistry in translucent molecular clouds.
\citet{GloverClark2012b} showed that the relatively small NL99 network reproduces the C$^{+}$, C, CO transition very well when compared to the more complex network of \citet{Glover2010}, whilst saving a factor of 3 in computational time.
We extend the NL99 network by differentiating between the $^{12}$C and $^{13}$C isotopes. The temperature-dependent chemical fractionation reaction \citep{Watson1976} and the differential shielding of CO isotopes from the interstellar UV radiation are also introduced to the chemical model.

In the solar metallicity case, we adopt elemental abundances (relative to hydrogen nuclei) of $x_{\textrm{He}} = 0.079$, $x_{^{12}\textrm{C}} = 1.4 \times 10^{-4}$, $x_{\textrm{O}} = 3.2 \times 10^{-4}$ and $x_{\textrm{M}} = 1 \times 10^{-7}$ for helium, carbon-12 isotope, oxygen and low ionisation potential metals (e.g. Na, Mg), respectively. We scale elemental abundances linearly with the adopted metallicity. The exception is the helium abundance, which is kept independent of metallicity.
Initially, the hydrogen is assumed to be fully molecular ($x_{\textrm{H}_2} = 0.5$), the oxygen and helium are atomic neutral, while the carbon and the metals are in singly ionised forms. This composition is inspired by the self-consistent cloud formation models of \citet{Clark2012b} and \citet{Smith2014}.
The initial $^{12}\textrm{C}/^{13}\textrm{C}$ isotopic ratio is set to 60, a value consistent with the measured average in the solar neighbourhood \citep{LucasLiszt1998}. For further details on the treatment of CO isotope chemistry we refer to \citet{Szucs2014}.

The model clouds are irradiated by an isotropic interstellar radiation field. The spectral shape of the ISRF is described by \citet{Draine1978} and \citet{Black1994} at ultraviolet and longer wavelengths, respectively. The fiducial radiative energy flux ($G_{0}$) is equivalent to 1.7 in units of the \citet{Habing1968} field, or $2.7\times10^{-3}~\textrm{erg}~\textrm{cm}^{-2}~\textrm{s}^{-1}$ integrated over the 91.2--240 nm wavelength range.
The H$_{2}$ and CO molecules are easily dissociated by the ultraviolet photons in optically thin gas \citep{DishoeckBlack1988}. As the H$_{2}$, CO and dust column densities build up at higher cloud depths, the molecules become shielded from the dissociating radiation. To account for the shielding effects we use H$_{2}$, $^{12}$CO, $^{13}$CO column density and visual extinction dependent scaling factors to adjust the photodissociation rates of these species. The adopted model for H$_{2}$ (self-)shielding is described in \citet{DraineBertoldi1996}. The factors of CO self-shielding and CO shielding by the Werner-band of H$_{2}$ molecules are taken in a tabulated form from \citet{Visser2009b}. The column densities were calculated using the {\sc TreeCol} algorithm \citep{Clark2012a}.

The SPH particle mass is kept consistent in both the $10^4\,\textrm{M}_{\odot}$ and the $10^5\,\textrm{M}_{\odot}$ simulation (at $0.05\, \textrm{M}_{\odot}$) by increasing the SPH particle number from $2\times10^6$ to $2\times10^7$. The factor of 10 increase in particle number results in more than a factor of 10 longer runtime.
A significant population of giant molecular clouds as massive as $10^6 - 10^7\, \textrm{M}_{\odot}$ is observed in the Milky Way and other galaxies in the Local Group \citep{Rosolowsky2005}. High resolution modelling of such massive clouds is very expensive in terms of computational time, and is therefore out of the scope of this paper.

For the further analysis, we select the last snapshot before the formation of the first sink particle. This choice is motivated by the fact that it takes time for the turbulence to wash away the artificial initial conditions (i.e. spherical symmetry, uniform density), and because stellar feedback is not included in the simulations (which would significantly affect the chemical and dynamical state of the system once protostars begin to form). The exceptions are models (g) and (h), the high virial parameter analogues of the fiducial model (d). In those runs the star formation is delayed, but for a consistent comparison we analyse them at the same time step as model (d).

\renewcommand{\thefigure}{\arabic{figure}}

\renewcommand{\thefigure}{\arabic{figure} (Cont.)}
\addtocounter{figure}{-1}

\begin{figure*}
\begin{center}
\includegraphics[trim=0 0 0 320,scale=0.64]{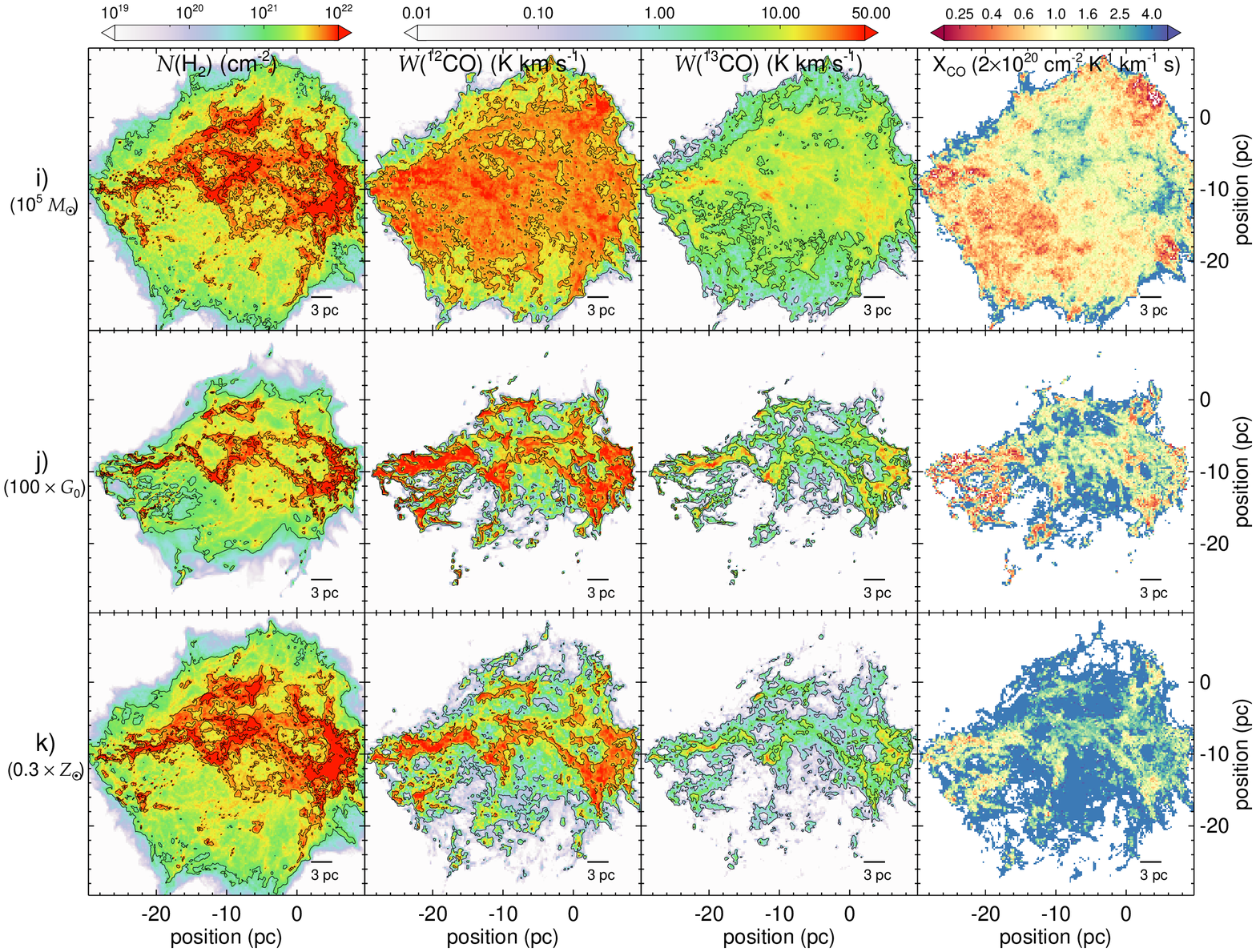}
\end{center}
\caption{H$_{2}$ column density map and the $^{12}$CO and $^{13}$CO integrated emission maps as well as the true $X_{\textrm{CO}}$ distribution for models (i) to (k). The complete set of models is available in the electronic-only material.}
\end{figure*}

\renewcommand{\thefigure}{\arabic{figure}}

\section{CO isotope emission maps} \label{sec:radtran}

We calculate position-position-velocity (PPV) emission maps in the $J=1\rightarrow0$ rotational transitions of $^{12}\textrm{CO}$  ($\lambda_{0,^{12}\textrm{CO}} = 2600.76\,\mu$m) and $^{13}\textrm{CO}$ ($\lambda_{0,^{13}\textrm{CO}} = 2720.41\,\mu$m). The spatial size of the maps is $16\,\textrm{pc}\times16\,\textrm{pc}$ with a linear resolution of 0.032 pc in the $M_{\textrm{cloud}} = 10^{4} \, \textrm{M}_{\odot}$ case. For the $M_{\textrm{cloud}} = 10^{5} \, \textrm{M}_{\odot}$ simulations the map size is about $40\,\textrm{pc} \times 40\,\textrm{pc}$ with a spatial resolution of 0.08 pc. The former resolution is comparable to single-dish sub-millimetre observations of low-mass, nearby molecular clouds, such as the Taurus cloud \citep{Pineda2010}, while the latter is from a factor of 3 to an order of magnitude better than resolution observed for distant, Galactic high-mass molecular cloud complexes, such as W43 \citep{Carlhoff2013} and $\eta$ Carinae \citep[][]{Yonekura2005}.
The emission is considered in a $\pm 6\,\textrm{km\,s}^{-1}$ velocity range around rest frame wavelengths of the transitions. The spectral resolution is set to $0.09\,\textrm{km\,s}^{-1}$. For the calculations we use the {\sc radmc-3d}\footnote{\protect\url{http://www.ita.uni-heidelberg.de/\~dullemond/software/radmc-3d/}} radiative transfer tool, described in \citet{Dullemond}. {\sc radmc-3d} is a grid based code, therefore the SPH data from the simulations are interpolated to a $512^{3}$ zone regular grid. The interpolation is performed using the {\sc splash}\footnote{\protect\url{http://users.monash.edu.au/\~dprice/splash/}} SPH data visualization tool \citep{Price2007}. We refer to \citet{Glover2015a} and \citet{Bertram2015c} for a resolution study. The choice of the SPH-to-grid interpolation scheme and its effect on the emission map are discussed in Appendix~\ref{appdx:restest} of this paper.

The critical hydrogen nuclei number density of the CO molecule $J=1\rightarrow0$ transition -- at which the collisional rate and the spontaneous de-excitation rate are equal -- is about $2200\,\textrm{cm}^{-3}$. Above this value the rotational energy levels are expected to be populated according to the thermal distribution (i.e. LTE holds). However, non-negligible CO emission already emerges where the number density reaches a few hundred particles per $\textrm{cm}^3$. Here, the CO excitation temperature is lower than the kinetic temperature and therefore the LTE approximation of radiative transfer is not applicable.
By volume, such regions dominate the simulation domains. For all simulations the majority of the volume (90 per cent of more) is below the CO critical density of $2200\,\textrm{cm}^{-3}$, thus a non-LTE radiative transfer approach is necessary.
Furthermore, in the optically thick regions the higher energy levels might be more populated than in LTE, due to  repeated excitation by trapped photons \citep[see Section 16.3.1 in][]{Wilson2009}.
To account for the sub-thermal excitation and photon trapping, we adopt the Large Velocity Gradient \citep[LVG,][]{Sobolev1957,Ossenkopf1997,Shetty2011a} method. In the LVG approximation, the global problem of radiation propagation is localised by correlating the escape probability of photons from a cell with the velocity gradient to the adjacent cells. \citet{Ossenkopf1997} showed that the approximation is reliable within 20\% even when the large gradient criterion does not strictly apply (e.g. in regions of in- or outflow).

Besides the number density distribution of the species in question and the gas temperature (a result from the hydrodynamic simulations), the properties (i.e. density and cross-section) of the collisional partners and the velocity vector field are also necessary for the LVG radiative transfer calculations. Due to their high abundance, hydrogen molecules are the most probable collisional partners for CO. We account for the two spin isomers of $\textrm{H}_{2}$ with an ortho-to-para number density ratio of 3, the ``hot'' formation ratio on grain surfaces. Due to chemical processing, the ratio can be orders of magnitude lower in the gas phase \citep[see e.g. ][]{Sipila2015}. We emphasise, however, that the exact choice of ortho-H$_{2}$/para-H$_{2}$ has only a few per cent effect on the CO emission \citep[see ][]{Szucs2014}. The collisional rates and the line properties are adopted from the Leiden Atomic and Molecular Database\footnote{\protect\url{http://home.strw.leidenuniv.nl/\~moldata/}} \citep{Schoier2005,Yang2010}. The resolved velocity field is taken from the hydrodynamic simulation. The velocity dispersion on sub-voxel scales may make a non-negligible contribution to the emission line width. To account for this, we assume that the velocity dispersion follows the observed scaling relation between size scale and line width (i.e. velocity dispersion), $v_{\textrm{unres}}=1.1\times(\Delta x / \textrm{pc})^{0.38}$ km s$^{-1}$ \citep{Larson1981,HeyerBrunt2004,Falgarone2009}, where $\Delta x$ is the linear size of a pixel in our case. The $10^{4}$ and the $10^{5} \, \textrm{M}_{\odot}$ models yield unresolved velocity dispersions of $0.29$ km s$^{-1}$ and $0.42$ km s$^{-1}$, respectively.

The maps are converted to brightness temperature ($[T_{\textrm{b}}] = \textrm{K}$) units for the sake of consistency with the observations.

\section{Methods for cloud mass estimation} \label{sec:massmethods}

There are several methods in the literature for inferring molecular cloud masses. Here we focus on those which are based on the measurement of CO emission. The most commonly used techniques include the measurement of CO isotope column density \citep[e.g.][]{Pineda2008,RomanDuval2010}, the virial analysis \citep[e.g.][]{MacLaren1988,Hughes2010} and the direct conversion of emission to $\textrm{H}_{2}$ column density with the $X_{\textrm{CO}}$-factor \citep[see the review of][]{Bolatto2013}. In the following three sub-sections, we describe these methods in greater details and apply them to the synthetic emission maps obtained from our simulations. We aim to test which methods work best under which conditions.

\subsection{Detection thresholds} \label{sec:detlim}

We restrict our analysis to regions which are above certain brightness temperature thresholds. The choice of the threshold levels is inspired by detection limits of real observations. We choose the $^{12}$CO and the $^{13}$CO detection limits to be $0.6\,\textrm{K}$ and $0.3\,\textrm{K}$, respectively. These are comparable to the single dish per channel $3\,\sigma_{\textrm{rms}}$ (root-mean-square noise) threshold of nearby \citep{Lee2014}, Galactic \citep{RomanDuval2010} and Large Magellanic Cloud \citep{Hughes2010} molecular cloud observations. The limits are applied consistently in each method. When the method requires the PPV cubes as input (i.e. the column density and the virial mass estimate), each PPV voxels below the threshold is omitted from the analysis. When the method requires $0$th moment (i.e. velocity channel integrated) maps, as in case of the CO-to-H$_{2}$ conversion factor method, then we calculate masked moment maps \citep[e.g.][]{Dame2001}: In a given line of sight, only channels with brightness temperatures higher than the detection limit are considered in the integration.
These approaches for applying detection limits on synthetic observations are, in effect, equivalent.

Note that some of the methods require only the $^{12}$CO emission as input. In these cases, all the PPV voxels with brightness temperatures higher than the $^{12}$CO detection limit are considered. Depending on the simulation parameters (i.e. metallicity and ISRF strength), a significant amount of molecular mass might reside in regions where the $^{12}$CO is brighter then the threshold, but the $^{13}$CO emission falls beneath it. This mass, in an ideal case, should be traced by the methods relying solely on $^{12}$CO emission, but it is ignored by definition in methods that rely on both isotopes. Therefore in our analysis, we distinguish the molecular mass in regions where both the $^{12}$CO and $^{13}$CO emission are above the threshold, from the molecular mass of regions where only the $^{12}$CO is required to be above the limit.
The methods are benchmarked taking this difference into account.

We neglect any additional uncertainties resulting from the intrinsic noise of the observational data and possible calibration errors.

\begin{table}
\caption{List of the tested methods. The abbreviations below are adopted in the reminder of this paper. See section~\ref{sec:massmethods} for the detailed description.}
\label{tab:methods}
\begin{center}
\begin{tabular}{ccc}
  \hline
     Abbreviation  & Method type & Reference  \\

  \hline

      \emph{$\textrm{W2009}_{\textrm{col}}$}   & column density  & \citet{Wilson2009} \\
      \emph{$\textrm{RD2010}_{\textrm{col}}$}  & column density  & \citet{RomanDuval2010} \\
      \emph{$\textrm{RL2006}_{\textrm{vir}}$}  & virial mass  & \citet{RosolowskyLeroy2006} \\
      \emph{$\textrm{ML1988}_{\textrm{vir}}$}  & virial mass  & \citet{MacLaren1988} \\
      \emph{$\textrm{GML2011}_{\textrm{XCO}}$} & $X_{\textrm{CO}}$-factor  & \citet{GloverMacLow2011} \\
      \emph{$\textrm{GAL}_{\textrm{XCO}}$}     & $X_{\textrm{CO}}$-factor  & e.g. \citet{Bolatto2013} \\
      \emph{$\textrm{W2010}_{\textrm{XCO}}$}   & $X_{\textrm{CO}}$-factor  & \citet{Wolfire2010} \\

 \hline
\end{tabular}
\end{center}
\end{table}

\subsection{Column density determination} \label{sec:coldens}
In this approach, first the column density of an optically thin molecular gas tracer (usually $^{13}\textrm{CO}$ or $\textrm{C}^{18}\textrm{O}$) is calculated. An abundance ratio between the observed species and $\textrm{H}_{2}$ is assumed (this implicitly contains a conversion to $^{12}$CO) to obtain the H$_{2}$ column density. This is then spatially integrated over the chosen column density contour level to give the total $\textrm{H}_{2}$ mass above the limit.

We start with the synthetic $^{13}\textrm{CO}$ emission maps derived from the simulations. The $^{13}\textrm{CO}$ column density, analogous to what we would measure from observations, is calculated following the concepts presented in \citet{Wilson2009} for linear molecules in LTE.
The method involves a set of additional assumptions, such as that all CO isotopes have a uniform excitation temperature along a line of sight and that their emission originates from the same volume. Additionally, the optical depth of the $J=1\rightarrow0$ transition is taken to be much larger than unity for $^{12}\textrm{CO}$ and few or less for $^{13}\textrm{CO}$. Furthermore, the only source of background radiation is the 2.7 K CMB.
It is known that some of these assumptions are invalid in realistic conditions or only hold over a limited column density range \citep{KennicuttEvans2012}. For instance, the $^{13}\textrm{CO}$ emission already becomes optically thick around a $^{13}$CO column density of $\textrm{few}\,\times\,10^{16}\,\textrm{cm}^{-2}$, and provides only a lower limit at higher columns. Similarly, the excitation temperature inferred from $^{12}$CO emission might differ considerably from the true excitation temperature \citep{Padoan2000,Molina2014}.
Despite these issues, many observational studies rely on this method \citep[e.g.][]{Goldsmith2008,Pineda2008,Pineda2010}.

To determine the column density of $^{13}\textrm{CO}$, first the ``effective'' excitation temperature along the lines of sight \citep[see][for a discussion of the choice of an extinction temperature measure]{Molina2014,Glover2015a} and the optical depth ($\tau$) of the emission line is estimated.
Generally, the intensity of an emission line is derived from the radiative transfer equation \citep[e.g.][]{Rybicki1986} and written as
\begin{equation}
I_{\textrm{line}} = (S - I_{0}) (1-\textrm{e}^{-\tau}),
\label{eq:intens}
\end{equation}
where $S$ denotes the source function, which is defined as the ratio of the emission and the absorption coefficients, and $I_{0}$ is the incident background intensity.

In the Rayleigh-Jeans regime, the brightness temperature ($T_{\textrm{b}}$) is strictly proportional to the kinetic temperature of the emitting gas. In fact, $T_{\textrm{b}}$ is defined as the temperature, which, when inserted into the Rayleigh-Jeans law, would give the measured intensity,
\begin{equation}
T_{\textrm{b}} = I_{\nu} \frac{c^2}{2 \nu^{2} k_{\textrm{B}}},
\label{eq:btmp}
\end{equation}
where $c$ is the speed of light in $\textrm{cm}\,\textrm{s}^{-1}$, $\nu$ is the transition frequency (115.271 GHz and 110.201 GHz for the lowest $J$ rotational transition of $^{12}$CO and $^{13}$CO, respectively) and $k_{\textrm{B}}$ is the Boltzmann constant in erg K$^{-1}$ unit. The brightness temperature and all other temperatures in the following equations are measured in Kelvin.
Equations \ref{eq:intens} and \ref{eq:btmp} can be combined to
\begin{equation}
T_{\textrm{b}} = T_{0} \left(\frac{1}{\textrm{e}^{T_{0}/T_{\textrm{ex}}} - 1} -  \frac{1}{\textrm{e}^{T_{0}/T_{\textrm{bg}}} - 1} \right) (1-\textrm{e}^{-\tau}),
\label{eq:tb}
\end{equation}
where $T_{0} = h\nu/k_{\textrm{B}}$, with $h$ denoting the Planck constant in CGS units. Assuming that the $^{12}$CO ($J=1\rightarrow0$) line is optically thick ($\tau \gg 1$), its excitation temperature is given by
\begin{equation}
  T_{\textrm{ex}}=5.5 ~\textrm{ln}\left(1+\frac{5.5}{T_{\textrm{b,peak}}^{12}+c_1}\right)^{-1},
\label{eq:tex}
\end{equation}
where the $T_{\textrm{b,peak}}^{12}$ is the $^{12}\textrm{CO}$ brightness temperature at the emission peak and $5.5\,\textrm{K} \equiv T_{0} = h\nu(^{12}\textrm{CO}) / k_{\textrm{B}}$.

We assume that the excitation temperatures of $^{12}$CO and $^{13}$CO are equal along each line of sight. Substituting the excitation temperature and the $^{13}$CO brightness temperature, $T_{\textrm{b}}^{13}(\nu)$, in equation~\ref{eq:tb}, we solve for
\begin{equation}
  \tau_{13}(\nu)=-\textrm{ln}\left[1-\frac{T_{\textrm{b}}^{13}(\nu)}{5.3} {\left\{\exp\left(\frac{5.3}{T_{\textrm{ex}}}-1\right)^{-1}-c_2\right\}^{-1}}\right],
\label{eq:tau}
\end{equation}
the $^{13}$CO ($J=1\rightarrow0$) optical depth, where $5.3\,\textrm{K} = h\nu(^{13}\textrm{CO}) / k_{\textrm{B}} $. The constants $c_1$ and $c_2$ correct for the background radiation. If the $2.7$ K CMB is the dominant source of the background radiation, then $c_1 = 0.82$ and $c_2 = 0.16$ \citep{Wilson2009}. Our synthetic emission maps, however, exclude any contribution from the background\footnote{In the hydrodynamic simulation we model the heating from a uniform interstellar radiation background that also accounts for the cosmic microwave background. In the radiative transfer post-processing only the emission from the molecular cloud is modelled.}, therefore both constants are set to zero.

The optical depth calculated in this way is an indicator for the column density of $^{13}$CO molecules in the lower $J$ state of the transition (in this case the ground state). To convert this to the total $^{13}$CO column density, we must sum over all energy levels of the molecule. Assuming that $T_{\textrm{ex}} = T_{\textrm{kin}}$ for all energy states and that the levels are populated according to the Boltzmann distribution, the population of all states can be estimated \citep[see equations 15.32 to 15.35 in][]{Wilson2009}. Thus, the column density of $^{13}\textrm{CO}$ is calculated according to
\begin{equation}
  N({^{13}\textrm{CO}})=3.0\times10^{14}\frac{ T_{\textrm{ex}}\int{\tau_{13}(v)dv}}{{1-\exp(-5.3/T_{\textrm{ex}})} },
\label{eq:n13co}
\end{equation}
where the beam (i.e. pixel) average column density is in the units of $\textrm{cm}^{-2}$ and the velocities ($v = 10^{-5}\,c\,(1-\nu / \nu(^{13}\textrm{CO}))$) are in $\textrm{km s}^{-1}$.
If the optical depth of the line is only a factor of a few larger than unity, then the numerator can be approximated as $T_{\textrm{ex}}\int{\tau_{13}(v)dv} = \tau_{13,0} / (1-\exp{(-\tau_{13,0})}) \int{T_{b}^{13}(v)dv}$, where $\tau_{13,0}$ is the optical depth of the line centre and $\int{T_{b}^{13}(v)dv}$ is the integrated line intensity. For $\tau_{13,0} > 2$, the expression is expected to overestimate the $^{13}$CO column density \citep[e.g.][]{Pineda2010}.
For the remainder of the paper, we call this method \emph{$\textrm{W2009}_{\textrm{col}}$} \citep[after][]{Wilson2009}.

\citet{RomanDuval2010} adopts a slightly different method (\emph{$\textrm{RD2010}_{\textrm{col}}$}); instead of calculating an effective excitation temperature along a line of sight, they calculate it for each velocity channel. Consequently, $T_{\textrm{ex}}$ becomes a function of velocity (i.e. frequency), and cannot be moved out from the integral. Note that, this method does not approximate the optical depth dependence, but solves the full integral.

The $^{13}$CO column density is converted to $^{12}$CO column density by multiplying it with the  $^{12}$CO/$^{13}$CO isotopic ratio. The direct determination of the ratio is difficult and restricted to ultraviolet and millimetre-wavelength absorption measurements \citep[][]{LisztLucas1998,Sheffer2007,Sonnentrucker2007}, which trace a lower column density range than those of typical molecular clouds.
In observational studies of MCs, often a single ratio is adopted for the whole cloud. This is chosen according to the average $^{12}$C/$^{13}$C ratio of the ISM within a few kpc of the Sun \citep{LangerPenzias1990,Wilson1999}. In practice, both the elemental and the $^{12}$CO/$^{13}$CO ratios are expected to vary within the same cloud, due to isotope selective photodissociation and chemical fractionation reactions \citep[see e.g.][]{Szucs2014}. The spatial variations and the deviation from the elemental ratio therefore contribute to the error of the CO column density based mass measurement method.

In the final step, the $^{12}$CO column density is converted to H$_{2}$ column density. For this, we assume that all the available carbon is incorporated in CO and that the gas is fully molecular ($x_{\textrm{H}_{2}} = 0.5$). In case of the solar metallicity model (c to j in Table~\ref{models}), this implies a $^{12}$CO to molecular hydrogen number density ratio, $n(^{12}\textrm{CO}) / n(\textrm{H}_{2}) = 2.8 \times 10^{-4}$. For the lower metallicity cases (a, b and k), the value scales proportionally to $Z$. With this choice of the $n(^{12}\textrm{CO}) / n(\textrm{H}_{2})$ ratio, the derived H$_{2}$ column density at the low end of the distribution is expected to systematically overestimate the true distribution. The reason for this is the efficient destruction of CO by photodissociation in the weakly shielded, low column density gas \citep{Visser2009b}, that decreases the ratio.
Nevertheless, when only CO observations are available for a given cloud, this approximation might be considered as the ``best guess''. For instance, \citet{Pineda2010} finds in the Taurus molecular cloud, when comparing H$_{2}$ column densities derived from CO isotope emission and visual extinction, that in regions where $^{12}$CO emission, but no $^{13}$CO  is observed (presumably translucent regions), the $n(^{12}\textrm{CO}) / n(\textrm{H}_{2})$ ratio varies strongly, while where both $^{12}$CO and $^{13}$CO emission are detected (likely dense, shielded regions), it can be approximated with a single value of $1.1 \times 10^{-4}$. More recently, \citet{Ripple2013} finds similar environment-dependent variations in the $^{13}$CO abundance in the Orion molecular cloud.

\subsection{Virial mass} \label{sec:virial}
The virial mass analysis is based on the measurement of the size and the average turbulent line width of the cloud. If these quantities are known and the radial density profile is given, then following \citet{Solomon1987}, the mass of the cloud under the assumptions of virial equilibrium and spherical symmetry can be calculated using the equation
\begin{equation}
M_{\textrm{vir}}\,=\,\frac{3(5-2\gamma)}{G(3-\gamma)}\,R_{\textrm{pc}}\,\Delta{v}^{2},
\label{eq:mvir1}
\end{equation}
where $\gamma$ is the exponent of the radial density distribution (i.e. $\rho(r)\,\propto\,r^{-\gamma}$), $G$ is the gravitational constant ($G \approx 1/232\,\textrm{M}_{\odot}^{-1}\,\textrm{pc}\,\textrm{km}^2\,\textrm{s}^{-2}$), $R_{\textrm{pc}}$ is the linear cloud size in pc and $\Delta{v}$ is the full width at half-maximum (FWHM) of the line in $\textrm{km s}^{-1}$, the result, $M_{\textrm{vir}}$ is given in solar masses (M$_{\odot}$).
Note that the assumption of spherical symmetry is a large simplification, since most clouds show highly filamentary structure \citep[][and references within]{Andre2014}.

We measure the cloud size and velocity dispersion in the $^{12}$CO position-position-velocity data cube using the moment-based method developed by \citet[][\emph{$\textrm{RL2006}_{\textrm{vir}}$} hereafter]{RosolowskyLeroy2006} and frequently adopted in observational studies \citep[e.g.][]{Hughes2010}. The cloud is defined by a brightness temperature iso-surface, such that all the connected pixels brighter than the limit $T_{\textrm{b,edge}}$ are associated with the cloud. We set the limit to be $T_{\textrm{b,edge}} = 0.6\,\textrm{K}$ (in the case of $^{12}$CO). The data cubes are then rotated such that the $x$ and $y$ axes are aligned with the major and minor axes of the cloud, respectively \citep[see equation 1 in][]{RosolowskyLeroy2006}. The root-mean-square cloud size is given by the geometric mean of the second spatial moments along the major and minor axes. The velocity dispersion is computed by taking the second moments along the velocity axis.
Assuming a Gaussian line profile, the velocity dispersion is related to the FWHM value via $\Delta{v}(T_{\textrm{b,edge}}) = \sqrt{8\,\ln(2)}\,\sigma_{\textrm{v}}(T_{\textrm{b,edge}})$, where $\sigma_{\textrm{v}}(T_{\textrm{b,edge}})$ is the velocity dispersion.
Taking the frequently assumed $\gamma=1$ radial density distribution exponent, equation~\ref{eq:mvir1} can be written as
\begin{equation}
M_{\textrm{vir}}(T_{\textrm{b,edge}})\,=\,1040\,\times\,R_{\textrm{pc}}(T_{\textrm{b,edge}})\,\times\,\sigma_{\textrm{v}}^{2}(T_{\textrm{b,edge}}),
\label{eq:mvir2}
\end{equation}
where the numerical coefficient accounts for the radial density profile, the conversion factor between line width and velocity dispersion and the gravitational constant \citep{MacLaren1988}. The result is again given in solar masses (M$_{\odot}$).

We use the {\sc cprops}\footnote{\protect\url{https://github.com/low-sky/cprops}} implementation of this recipe. An additional feature of {\sc cprops} is the possibility of extrapolating the cloud properties (size and velocity dispersion) to the $T_{\textrm{b,edge}} = 0\,\textrm{K}$ contour level (i.e. to correct for emission falling below the detection limit). We do not use this option for the sake of consistency with the other methods.

\citet[][\emph{$\textrm{ML1988}_{\textrm{vir}}$}]{MacLaren1988} proposed a slightly different analysis by suggesting the use of both $^{12}\textrm{CO}$ and $^{13}\textrm{CO}$ emission. They argue that the virial theorem gives a reliable result only if the considered $\Delta{v}$ describes the average line-width over the whole cloud, including the central core regions. The typically optically thick $^{12}\textrm{CO}$ emission provides information only from the cloud surface (where the optical depth is about 1), while $^{13}\textrm{CO}$ remains optically thin until larger depths and probes the velocity structure deeper in the cloud. On the other hand, due to its preferential photo-dissociation, $^{13}\textrm{CO}$ has a larger formation threshold in $A_{\textrm{V}}$ than $^{12}\textrm{CO}$, and cloud size estimates based on the former would underestimate the true value.
In our implementation of the \emph{$\textrm{ML1988}_{\textrm{vir}}$} method, we also use the {\sc cprops} algorithm to calculate the cloud size and the velocity dispersion, but we obtain the former from the $^{12}$CO emission, and the latter from the $^{13}$CO emission (with $T_{\textrm{b,edge}} = 0.3 \, K$).

The virial mass of the cloud is a measure of the total cloud mass, thus the helium and atomic hydrogen (H{\sc i}) content should be subtracted from it to arrive at the H$_{2}$ mass. The helium mass fraction is 0.24 in each simulation. This is a constant contribution, and we simply subtract it from the virial mass estimates. The contribution of H{\sc i} to the total- and CO-bright mass budget is more model dependent. In the low metallicity or high ISRF cases, the H{\sc i} component of the total cloud mass might amount up to 10 per cent. In the case of the Milky Way-like model the contribution is 6 per cent. Furthermore, the distribution of H{\sc i} strongly varies in space. In CO-bright regions (dense gas) it only contributes up to 0.5 per cent of the total mass, even when the metallicity is low or the radiation field is strong. In these regions the gas is mostly molecular, and hydrogen is locked in H$_2$. Hence, we do not subtract the H{\sc i} mass from the virial estimate, due to its negligible contribution in the CO-bright gas.

\begin{figure}
\begin{center}
\includegraphics[trim=20 0 0 0,scale=0.51]{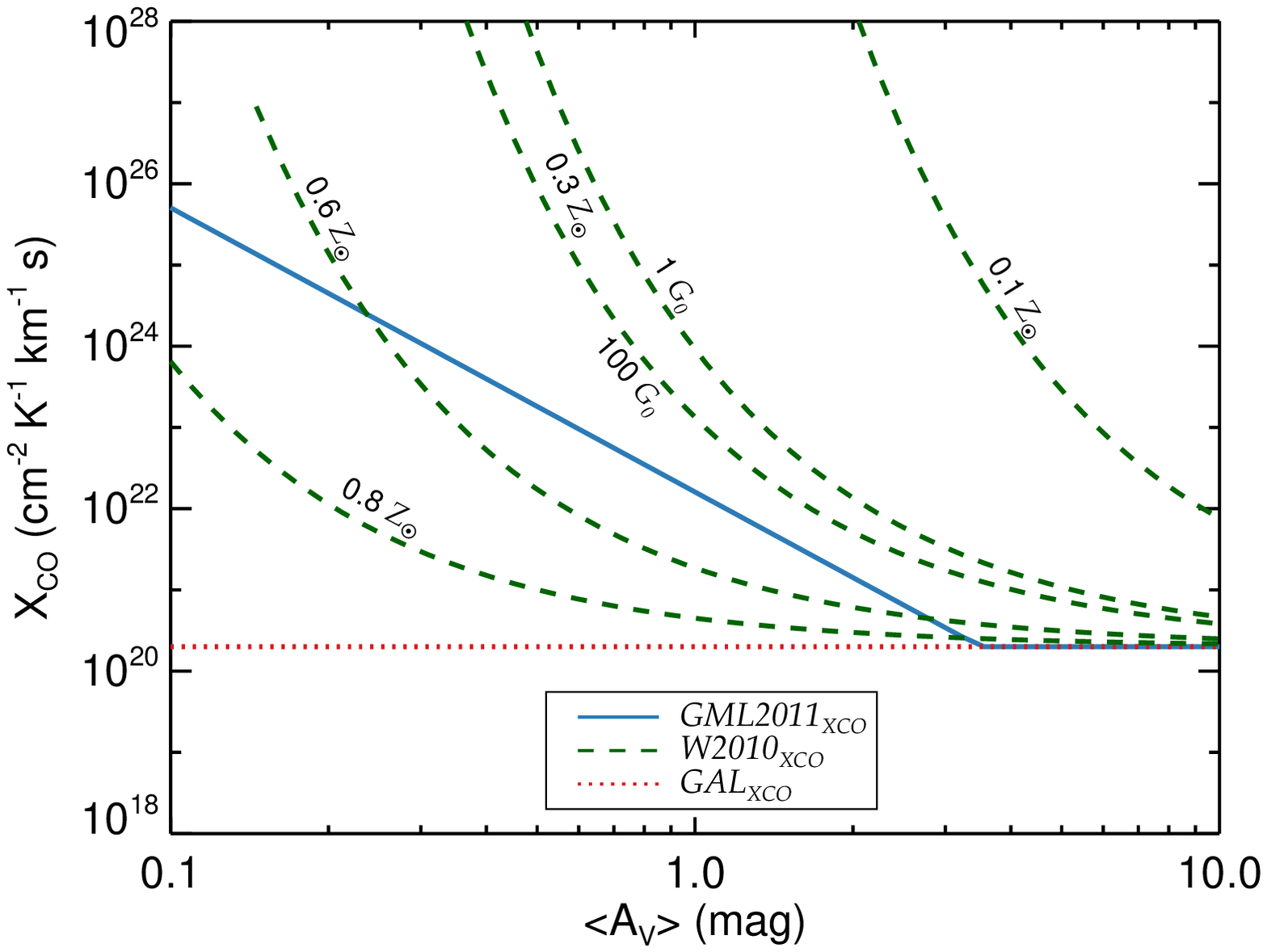}
\end{center}
\caption{Comparison of prescriptions for the mean visual extinction and metallicity dependent $X_{\textrm{CO}}$-factors. In the case of the \emph{$\textrm{GML2011}_{\textrm{Xco}}$} method, the metallicity dependence enters implicitly, via the scaling of the dust-to-gas ratio with $Z$. The \emph{$\textrm{W2010}_{\textrm{Xco}}$} method accounts for the metallicity and the ISRF dependence explicitly. In this case, $X_{\textrm{CO}}$ strongly depends on the metallicity, while the ISRF dependence is much weaker. At high $\langle A_{\textrm{V}} \rangle$ all models converge by construction to the Galactic value, while at low mean ${\textrm{A}_{\textrm{V}}}$ values there is a wide distribution of possible conversion factors. The \emph{$\textrm{GML2011}_{\textrm{X}_{\textrm{CO}}}$} empirical fit is roughly consistent with the 0.3 and 0.6 $Z_{\odot}$ metallicity \emph{$\textrm{W2010}_{\textrm{X}_{\textrm{CO}}}$} curves.}
\label{fig:xco}
\end{figure}

\subsection{\texorpdfstring{The CO-to-H$_{2}$ conversion factor}
                   {The CO-to-H2 conversion factor}} \label{sec:xfactor}

In the case of nearby molecular clouds, the H$_{2}$ column density can be estimated indirectly by measuring the total column density of hydrogen (using e.g. dust extinction, dust emission, or the diffuse $\gamma$-ray flux) and subtracting the column density of atomic hydrogen (measured by the H{\sc i} 21~cm hyperfine emission line).
A series of studies \citep[e.g.][]{Sanders1984,Strong1996,Hunter1997} found that the H$_{2}$ column density calculated this way is proportional to the velocity-integrated intensity of the $^{12}$CO ($J=1\rightarrow0$) line ($W_{\textrm{CO}} = \int T^{12}_{\textrm{B}}(v)\, dv$, in units of K km s$^{-1}$). The proportionality is often expressed with the so-called $X_{\textrm{CO}}$-factor:
\begin{equation}
  X_{\textrm{CO}} \equiv \frac{N_{\textrm{H}_2}} {W_{\textrm{CO}}}.
\end{equation}
A number of independent studies \citep[see the review of][]{Bolatto2013} find a Galactic mean value,
\begin{equation}
 X_{\textrm{CO},0} = 2 \times 10^{20}\, \textrm{cm}^{-2}\,\textrm{K}^{-1}\,\textrm{km}^{-1}\,\textrm{s},
\end{equation}
with 30\% uncertainty within the Milky Way disk environment (i.e. for $Z = Z_0$, $G = G_0$).

The $X_{\textrm{CO}}$-factor is, however, expected to vary on small scales and to correlate with the cloud properties and the (galactic) environment\footnote{See \citet{Lee2014} for an observational example in the Perseus cloud and \citet{Shetty2011a,Shetty2011b,Clark2015} for a theoretical investigation.}. In general, the variation in the conversion factor is due to the combined effects of changes in the chemical and excitation properties of the cloud with scale, environment and/or composition.
In low metallicity environments, such as dwarf irregular galaxies, the cloud average $X_{\textrm{CO}}$ factor might increase drastically, by orders of magnitude. As an example, \citet{Bolatto2011} find about 2 orders of magnitude increase in the Small Magellanic Cloud, compared to the Milky Way value \citep[see also][]{Tacconi2008}. The reasons could be traced back to both the lower elemental abundance of carbon and oxygen available for CO production and the reduced number density of dust particles that contribute to the shielding of molecules from the UV radiation.
The H$_{2}$ and CO molecules are formed through qualitatively different chemical processes and the efficiency of their destruction by the ISRF differs. H$_{2}$ mainly forms on dust grains and its abundance depends primarily on the available time \citep[e.g. in conditions typical to the turbulent Milky Way clouds, the H$_{2}$ formation timescale is a few Myr; see e.g. ][]{GloverMacLow2007b}. CO, on the other hand, forms primarily in the gas phase via the relatively fast ion-neutral and the somewhat slower neutral-neutral reactions and due to its less efficient shielding it gets destroyed deeper into the cloud than H$_{2}$, thus its abundance is set mainly by the shielding available \citep[for further discussion we refer to ][]{KlessenGlover2016}.
With decreasing metallicity, the CO shielding decreases, therefore the CO emitting zone shrinks to enclose only the highest (column) density regions of the clouds, while the H$_{2}$ rich gas can still remain extended. This results in an increasing cloud average $X_{\textrm{CO}}$ factor \citep[see, e.g. ][ and references therein]{Bolatto2013}.

Similarly, when the incident ISRF is strong, e.g.\ in the Galactic Centre or in a starburst environment, the CO-abundant zone retreats to higher total column densities. By itself, this effect will tend to increase $X_{\rm CO}$. However, the stronger irradiation also increases the gas temperature, particularly if it is accompanied by an increased cosmic-ray ionisation rate. This in turn increases the brightness of the CO line, and tends to decrease $X_{\textrm{CO}}$. Which of these effects dominates is not clear {\it a priori}, and it is plausible that in some circumstances they may cancel almost entirely, resulting in an $X_{\textrm{CO}}$ close to the Galactic value \citep{Liszt2010}.

This issue has recently been investigated numerically by \citet{Clark2015}. They study the effects on $X_{\rm CO}$ of increasing the strength of the ISRF and the size of the cosmic ray ionisation rate from the canonical local values to values a hundred times larger. They find that in small clouds with densities comparable to local MCs, the effect on $X_{\textrm{CO}}$ is rather small, while in larger clouds, there is a pronounced increase in $X_{\rm CO}$ as the radiation field strength and cosmic ray ionisation rate are increased, although even in this case the dependence is strongly sub-linear. However, they also show that if the mean cloud density and velocity dispersion are increased at the same time (as is plausible for clouds in extreme environments), then this can completely offset the effect of the harsher environment on $X_{\textrm{CO}}$.

The numerical simulations of \citet{GloverMacLow2011} and the analytical calculations of \citet{Wolfire2010}, formulated in \citet{Bolatto2013}, offer theoretical calibrations for the environment dependence of the cloud average $X_{\textrm{CO}}$ factor. \citet{GloverMacLow2011} perform turbulent, three-dimensional magneto-hydrodynamic simulations with self-consistent chemistry and thermal balance, while neglecting self-gravity. They explore solar and sub-solar metallicities and various mean cloud densities (from $n_{0} = 30$ to $1000\,\textrm{cm}^{-3}$) and determine the cloud average $X_{\textrm{CO}}$ factor as a function of the cloud-average visual extinction ($\langle {A}_{\textrm{V}} \rangle$). At low $\langle {A}_{\textrm{V}} \rangle$ the conversion factor approximately follows a power law, while above 3.5 mag cloud average visual extinction, the $X_{\textrm{CO}}$-factor converges to the Galactic disk value. We adopt the following functional form of their result
\begin{equation}
X_{\textrm{CO}} = \begin{cases} {X_{\textrm{CO},0}  { \left(\frac{\langle A_{V} \rangle }{3.5}\right)}^{-3.5} } & \mbox{if }  \langle A_{V} \rangle < 3.5\,\textrm{mag}\\
      X_{\textrm{CO},0} \qquad\qquad\quad  & \mbox{if } \langle A_{V} \rangle \geq 3.5\,\textrm{mag}. \end{cases}
\label{eq:gloverxco}
\end{equation}
In the remainder of this paper we reference the H$_{2}$ mass calculated with the $X_{\textrm{CO}}$ value derived this way as \emph{GML2011$_{XCO}$}.

\citet{Bolatto2013} determine the metallicity dependence of the conversion factor following the analytical argument of \citet[][hereafter \emph{W2010$_{XCO}$}]{Wolfire2010}. They assume a spherically symmetric molecular cloud with a radial density profile of $\rho \propto r^{-1}$ and calculate the difference in visual extinction between the CO-abundant (i.e. the cloud depth where the $^{12}$CO becomes optically thick) and the H$_{2}$-dominated layers (i.e. the depth where the gas is half molecular) of the cloud. This is given by
\begin{equation}
\Delta A_{\textrm{v}} = 0.53 - 0.045 \times \log_{10}\left({\frac{G}{n}}\right) - 0.097 \log_{10}(Z),
\end{equation}
where $n$ is the mean gas density, $G$ and $Z$ are
the ISRF strength and the metallicity in the units of $G_{0}$ and $Z_{\odot}$, respectively. The $X_{\textrm{CO}}$ factor is expected to be proportional to the mass ratio of the CO-bright and the total molecular gas masses, which is written as $X_{\textrm{CO}} \propto M(R(\textrm{H}_{2})) / M(R(\textrm{CO})) = \exp(-4 \Delta A_{\textrm{V}} / {\langle A_{\textrm{V}} \rangle} )$. The metallicity dependence is introduced through the visual extinction, by assuming that the dust-to-gas ratio decreases with decreasing metallicity \citep{Draine2007}, thus $\langle A_{\textrm{V}} \rangle (Z) = \langle A_{\textrm{V}} \rangle (Z_{\odot}) \times Z$. The metallicity dependence of $X_{\textrm{CO}}$ is then derived from the ratio $M(R(\textrm{H}_{2})) / M(R(\textrm{CO}))$ at metallicity $Z$ and the solar value ($Z_{\odot}$),
\begin{equation}
X_{\textrm{CO}} =  X_{\textrm{CO},0}  \exp{\left( \frac{4 \Delta A_{V}}{\langle A_{\textrm{V}} \rangle }\right)}\,\exp{\left( \frac{-4 \Delta A_{\textrm{V}}}{\langle A_{\textrm{V}} \rangle / Z } \right)}.
\label{eq:wolfirexco}
\end{equation}
In the first exponent, $\langle A_{\textrm{V}} \rangle$ denotes the cloud average visual extinction measured at metallicity $Z$. In the second exponent $\langle A_{\textrm{V}} \rangle / Z$ stands for the average visual extinction that could be measured for the same cloud, if its metallicity were $Z_{\odot}$.

\begin{figure}
\begin{center}
\includegraphics[trim=90 176 0 20, scale=0.64]{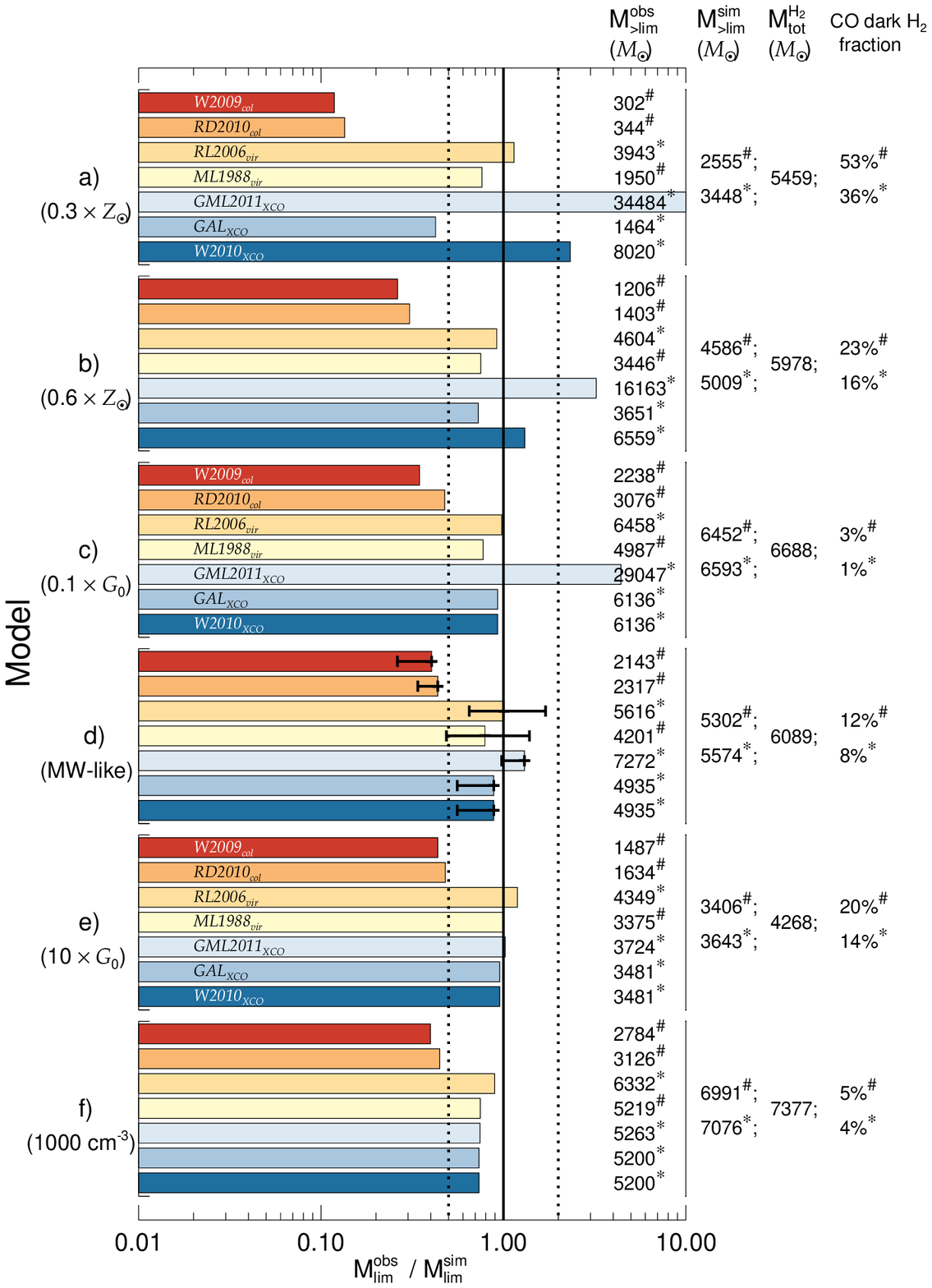}
\end{center}
\caption{Comparison of cloud masses derived using different methods. The horizontal axis shows the ratio of the measured H$_{2}$ mass and the true H$_{2}$ mass above the detection limit. The latter is known from the simulation. The numerical values of the H$_{2}$ mass and its estimates can be read from the right side of the figure. The total H$_{2}$ content of the clouds and their CO-dark H$_{2}$ fraction are also shown.
The solid vertical line shows the one-to-one ratio of the estimate and the true H$_{2}$ mass, while the dotted vertical lines indicate a factor of 2 deviation from it.
The error bars in case of d) represent the variation due to different viewing angles and cloud realisation (see Appendix~\ref{appdx:viewangle}).
The ``\#'' and ``$*$'' symbols indicate whether the method takes both the $^{12}$CO and $^{13}$CO detection limits into account, or only the $^{12}$CO threshold, respectively (see Section~\ref{sec:detlim}). The molecular mass above the limit and the CO-dark fraction is also indicated in the two cases.}
\label{fig:masscomp}
\end{figure}

\renewcommand{\thefigure}{\arabic{figure} (Cont.)}
\addtocounter{figure}{-1}

\begin{figure}
\begin{center}
\includegraphics[trim=90 250 0 20, scale=0.64]{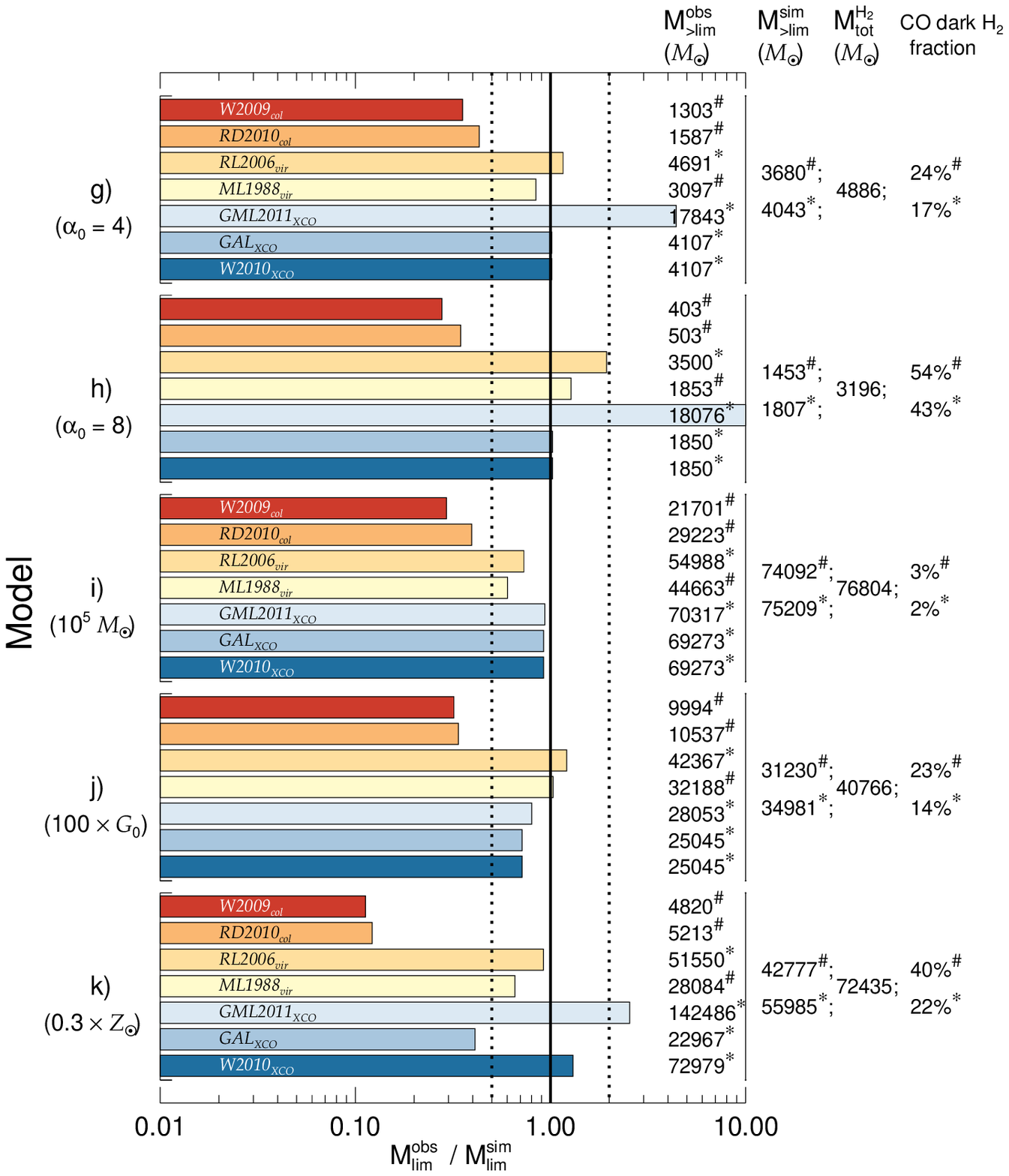}
\end{center}
\caption{Comparison of cloud masses derived using different methods from simulations (g) to (k). The ``\#'' and ``$*$'' symbols indicate whether the method takes both the $^{12}$CO and $^{13}$CO detection limits into account, or only the $^{12}$CO limit, respectively (see Section~\ref{sec:detlim}). The molecular mass above the limit and the CO-dark fraction is also indicated in the two cases.}
\end{figure}

\renewcommand{\thefigure}{\arabic{figure}}

Fig.~\ref{fig:xco} compares the $X_{\textrm{CO}}$-factors and their respective dependencies in the physical conditions, according to the \emph{GML2011$_{XCO}$} and the \emph{W2010$_{XCO}$} methods. For comparison we also show the Galactic $X_{\textrm{CO}}$ (\emph{GAL$_{XCO}$}) on the figure.

The total H$_{2}$ mass of the cloud is calculated by the following prescription. First, we integrate the $^{12}$CO ($J=1\rightarrow0$) PPV cubes along the velocity dimension to obtain the two-dimensional $W_{\textrm{CO}}$ map. The PPV brightness temperatures, which are below the $3 \,\sigma_{^{12}\textrm{CO}} = $ 0.6 K detection limit (see section~\ref{sec:detlim}) are omitted from the integration. Then the $W_{\textrm{CO}}(x,y) = 0\,\textrm{K km s}^{-1}$ pixels are rejected from the velocity-integrated map, and the arithmetic mean intensity value of the remaining pixels is calculated. With the latter step, the effects of small scale $X_{\textrm{CO}}$ variations are lessened. The cloud average visual extinction is the mean of the true visual extinction in the lines of sight (i.e. pixels) used to calculate the mean intensity. The true visual extinction is determined from the hydrogen nuclei column density via $A_{\textrm{V}} = 5.348 \times 10^{-22} N_{\textrm{H}}$ for dust-to-gas ratio of 1/100 \citep{Bohlin1978,DraineBertoldi1996}. The total H$_2$ mass estimate is then the product of the mean intensity, the adopted $X_{\textrm{CO}}$-factor, the surface area over which the average is taken and the H$_{2}$ molecule mass.

\section{Results} \label{sec:result}

Fig.~\ref{fig:masscomp} summarises the H$_{2}$ masses estimated using the methods described above for all of the simulations. In addition to the molecular, i.e. H$_{2}$, mass estimate (M$^{\textrm{obs}}_{>\textrm{lim}}$), it also lists the true molecular mass above the CO brightness temperature threshold M$^{\textrm{sim}}_{>\textrm{lim}}$ (CO-bright H$_2$ mass hereafter), the total H$_{2}$ mass (M$^{\textrm{H}_{2}}_{\textrm{tot}}$, i.e. the sum of CO-bright and CO-dark) and the true CO-dark molecular gas fraction (i.e. the ratio of H$_{2}$ mass below the detection limit and the total H$_{2}$ mass).

The CO-bright H$_{2}$ mass is defined as the integral of the true H$_{2}$ column density over the position-position surface, where, at least in a single velocity channel, the $^{12}$CO and $^{13}$CO brightness are both greater or equal to the respective 0.6 K and 0.3 K detection limits (marked by ``\#''), or where at least the $^{12}$CO brightness is above the threshold (marked by ``*''). For methods that take both the $^{12}$CO and $^{13}$CO emission into account (e.g. \emph{RD2010$_{col}$}), we use the former definition of the CO-bright mass, since these methods cannot be applied to lines of sight where $^{13}$CO is not detected. On the other hand, for methods that only make use of the $^{12}$CO emission (e.g. the methods based on the $X_{\textrm{CO}}$-factor), we use the latter definition.
The total H$_2$ mass is given by the integral of the true H$_2$ column density over the position-position surface, where $N(\textrm{H}_2$) is grater than zero.

In case of the fiducial model (d), the emission and the mass estimates are also calculated at different viewing angles. Additionally, we consider a second Milky Way-like cloud by running a hydrodynamic simulation with identical parameters but with a different random seed for the initial turbulent velocity field. The horizontal error bars represent the range that the estimates cover in these realisations. See Appendix~\ref{appdx:viewangle} for a detailed comparison.

We emphasise the difference between the CO-bright H$_2$ mass and the total H$_2$ mass of the cloud. In case of high metallicity and moderate ISRF strengths, the CO-bright and total H$_2$ masses are very similar (see models c, d, f, i). However, in case of sub-solar metallicities and/or strong ISRFs (a, b, e, j, k), or when the virial parameter is high (g, h), the difference is much larger, up to 54 per cent, due to the enhanced CO-dark molecular mass fraction.
We also find a factor of 1.4 - 1.5 difference in our estimate of the CO-dark H$_2$ fraction, depending upon whether we require both the $^{12}$CO and $^{13}$CO brightness temperatures to be above the detection limit, or only the $^{12}$CO brightness temperature. In other words, the mass of molecular gas that is completely CO-dark is roughly twice the size of the ``diffuse'' molecular component that is dark in $^{13}$CO but bright in $^{12}$CO \citep[c.f.][]{RomanDuval2016}.
In the following, we compare the mass estimates to the CO-\emph{bright} H$_2$ mass of the clouds (section~\ref{sec:res:co-bright}). Then we discuss the environment dependence of the CO-dark molecular mass fraction and evaluate the methods in terms of their ability to recover the \emph{total} H$_2$ mass (section~\ref{sec:res:co-dark}). Finally, we compare the virial parameter of the cloud inferred from the observed virial mass to the virial state of the CO-bright regions and the overall molecular cloud (section~\ref{sec:res:vir}).

\subsection{CO-bright H$_2$ mass} \label{sec:res:co-bright}

The LTE column density measurement methods always underestimate the CO-bright H$_2$ mass. In the best cases (c, d and e) the difference is about factor of 2, while in the worst cases (a and k) it is an order of magnitude.
We find that the virial and $X_{\textrm{CO}}$ masses are ``good'' indicators, i.e. within a factor of $2$, of the true cloud mass, unless the metallicity is low ($< 0.6 \times Z_{\odot}$). The exceptions is the \emph{GML2011$_{XCO}$} method, which yields about a factor of 10 error in the lowest metallicity cases or when the virial parameter is high.

The CO isotope column density measurement based methods (\emph{W2009$_{col}$} and \emph{RD2010$_{col}$}) are poor indicators of M$^{\textrm{sim}}_{>\textrm{lim}}$. They tend to perform similarly, with the largest discrepancy of $\sim30\%$, and in all cases the \emph{RD2010$_{col}$} provides a better estimate. These methods work best for solar metallicity and $G = 1, 10 \times G_{0}$ conditions, but even then they underestimate the CO-bright (and total) H$_{2}$ mass severely. Their mass estimate is largely insensitive to the virial parameter change (d, g and h).

The virial analysis based methods (\emph{RL2006$_{vir}$} and \emph{ML1988$_{vir}$}) seem to work the most consistently for each model. Surprisingly, this is the case even when the virial parameter is supercritical, thus the kinetic energy dominates over the potential energy and the velocity dispersion is not a direct indicator of the cloud mass. We discuss the reason for the good match in detail in section~\ref{sec:disvir}. The purely $^{12}$CO emission based \emph{RL2006$_{vir}$} method performs better in all cases, except when the ISRF strength or the virial parameter is high. In those cases it overestimates the H$_2$ mass, while the \emph{ML1988$_{vir}$} method, which consistently yields about $25\%$ lower masses, provides a more precise estimate.

The CO-to-H$_{2}$ conversion factor methods exhibit a large variation in the reliability of their mass estimates. The determining quantity in the values of the $X_{\textrm{CO}}$ factor in \emph{GML2011$_{XCO}$} and \emph{W2010$_{XCO}$} is the metallicity. The dependence is implicit in the first case, through the metallicity dependent visual extinction value, and explicit in the latter. The \emph{GML2011$_{XCO}$} method tend to perform the poorest in recovering M$^{\textrm{sim}}_{>\textrm{lim}}$ when metallicity is low (see models a, b and j) or when $\alpha_{\textrm{vir}}$ is high. The \emph{W2010$_{XCO}$} method overestimates the H$_2$ mass in the low metallicity cases by a similar factor (about $2$) as the \emph{GAL$_{XCO}$} underestimates it. The dependence on the virial parameter is due to the reduction of cloud average $A_{\textrm{V}}$ (see panels d, g, h in Fig.~\ref{fig:xcoall}).

The \emph{GAL$_{XCO}$} method provides an estimate consistent with the CO-bright H$_{2}$ mass within a factor of $2-3$.
The Galactic $X_{\textrm{CO}}$ value works the best at solar metallicity and it is only weakly sensitive to the ISRF strength. In these cases the CO-to-H$_{2}$ conversion factor does not vary strongly, and it is close to the Galactic value in the great majority of pixels (see panels e and j in Fig.~\ref{fig:maps} and Fig.~\ref{fig:xcoall}).

\subsection{CO-dark gas fraction and the total H$_2$ mass} \label{sec:res:co-dark}

The CO-dark (sometime called ``CO-faint'', in general H$_{2}$ gas not traced by CO emission) molecular gas fraction and its environmental dependence is a major focus of research in both the Galactic and extragalactic ISM studies. Dust emission based analysis of nearby molecular clouds finds that the CO-dark H$_{2}$ fraction is about $\sim30\%$ \citep{Planck2011}, and hints at a metallicity dependence.
The individual cloud scale PDR models presented in \citet{Wolfire2010} and the galactic scale hydrodynamic simulations of \citet{Smith2014} in a Milky Way-like setup find this fraction to be $30\%$ and $42\%$ respectively. The former also finds that the fraction is insensitive to cloud properties and environment.
We note that there are two main components of the CO-dark molecular gas: translucent clouds with low total visual extinction and the translucent envelopes of dense clouds. The models of \citet{Wolfire2010} are informative on the envelopes of dense clouds, while the \citet{Smith2014} models take both components into account.
The simulations presented here are more similar to the individual cloud models of \citet{Wolfire2010}. We probably underestimate the CO-dark molecular gas fraction, due to the fact that the simulations are initialised with isolated, homogeneous density spheres. Nevertheless, the models may provide insight into systematic changes with metallicity, ISRF strength and the adopted CO-based mass measurement method.

We find that the CO-dark gas fraction depends strongly on the cloud environment, metallicity and on the imposed detection threshold. When both the $^{12}$CO and $^{13}$CO thresholds are considered, the dark mass fraction is the largest in the low metallicity or high $\alpha_{\textrm{vir}}$ simulations (a and k, $Z = 0.3 \times Z_{\odot}$, h), where it is about or higher than 40 per cent. As the metallicity doubles (b), the fraction decreases by about a factor of 2 to $23\%$. At solar metallicity (d), it halves again to about $12\%$. The ISRF strength increase affects the fraction similarly to the decreasing metallicity. These changes in the amount of gas traced by CO can be followed on the corresponding panels of Fig.~\ref{fig:maps}.
When only the $^{12}$CO threshold is imposed, the CO-dark mass fraction is lower, by about 30-40 per cent.
A similar trend seems to apply for the $10^{5} \textrm{M}_{\odot}$ cloud (i, j and k); the CO-dark gas fraction increases by more than a factor of 10 from (i) to (k) as metallicity decreases.

\subsection{Observed virial parameter} \label{sec:res:vir}

\begin{table}
\caption{Measured and true virial parameters. The former, $\alpha_{\textrm{vir,measured}}$ is calculated as the ratio of the virial mass (eq.~\ref{eq:mvir2}) and the true molecular mass above the CO brightness limit (CO-bright H$_2$ mass) or the virial mass and the true total H$_2$ mass of the cloud (total H$_2$ mass). The true virial parameter is calculated using the physical quantities in the hydrodynamic simulation. The CO-bright true virial parameter refers to the virial parameter of the gas with CO abundance, $x_{\textrm{CO}} \geq 10^{-5}$. The total true virial parameter includes the complete cloud, these values are also listed in Table.~\ref{models}. }
\label{tab:viralpha}
\begin{center}
\begin{tabular}{ccccc}
  \hline
     Model  & \multicolumn{2}{c}{$\alpha_{\textrm{vir}}$ CO-bright} & \multicolumn{2}{c}{ $\alpha_{\textrm{vir}}$ Total} \\
        & measured & true & measured & true \\

  \hline
      a & 1.20 & 0.69 & 0.54 & 1.08 \\
      b & 0.93 & 0.88 & 0.71 & 1.10 \\
      c & 0.98 & 0.98 & 0.95 & 1.04 \\
      d & 1.02 & 0.92 & 0.88 & 1.04 \\
      e & 1.21 & 0.93 & 0.95 & 1.07 \\
      f & 0.90 & 1.08 & 0.85 & 1.51 \\
      g & 1.18 & 1.73 & 0.87 & 2.52 \\
      h & 2.00 & 2.96 & 0.88 & 7.95 \\
      i & 0.73 & 0.87 & 0.71 & 0.90 \\
      j & 1.20 & 0.73 & 0.93 & 0.89 \\
      k & 0.89 & 0.72 & 0.54 & 0.88 \\
 \hline

\end{tabular}
\end{center}
\end{table}

The fact that the virial mass seems to be a good measure of the true molecular mass above the CO detection limit suggests that both the observationally-inferred and the actual virial parameters of the CO-bright portion of the cloud must be close to unity. It worth examining, however, to what extent the observed value correlates with the overall virial state of the cloud.  In Table~\ref{tab:viralpha}, we list two different values for the observationally-inferred or ``measured'' virial parameter for each of our model clouds. In both cases, we compute the virial mass of the cloud using Eq.~\ref{eq:mvir2}. In the first case, we then compute a virial parameter by dividing this by the H$_{2}$ mass of the CO-bright regions of the cloud (defined here as those parts of the cloud where the CO fractional abundance exceeds $10^{-5}$). These values are denoted in the Table as $\alpha_{\textrm{vir}}$ CO-bright. In the second case, we instead divide $M_{\textrm{vir}}$ by the total H$_{2}$ mass. We compare these values with the ``true'' virial parameters of either the CO-bright molecular gas or the entirety of the molecular gas, computed using the output of our hydrodynamical simulations.

In case of the $\alpha_{\textrm{vir,0}} = 2$ simulations (a to f and i to j), we find observed virial parameters around unity with a scatter of roughly 20 per cent. The observed virial parameters of the high $\alpha_{\textrm{vir,0}}$ models are systematically underestimated, by 50 per cent when compared to the virial parameter of the CO-bright gas, and up to a factor of 4 when compared to the overall virial parameter of the cloud (see section~\ref{sec:disvir}).

Generally, the observed virial parameters are in good agreement with the virial parameter of the CO bright gas, when the true virial parameter is low, and underestimate the true value when it is much larger than unity. The overall virial parameter of the cloud is always underestimated. This is due to the underestimated virial mass of the cloud. We point out, that this implies that molecular clouds which are observed in CO emission to be subvirial, might in fact be stable to gravitational collapse or even expand on large scales.

\section{Further discussion}  \label{sec:dis}

We find that the CO isotope column density estimation methods seriously underestimate the true molecular cloud mass, while the other methods are accurate within a factor of $2$ in conditions similar to those in the Milky Way. But why do the former methods perform so poorly and why do the latter work reasonably well? How good are the column density estimate methods at recovering the true H$_{2}$ column density distribution? Why do the virial analysis methods work so well in the whole studied parameter range? What is the problem with the $A_{\textrm{V}}$ dependent $X_{\textrm{CO}}$-factors, and how does the CO-to-H$_{2}$ conversion factor change with physical conditions on the sub-parsec scale? In the following section we discuss these questions.

\subsection{\texorpdfstring{Inferring H$_{2}$ column density from CO isotope emission}
                   {Inferring H2 column density from CO isotope emission} } \label{sec:discol}

In the previous section we showed that the molecular mass derived by inferring the H$_2$ column density from the CO emission significantly underestimates the cloud mass in all cases. Consequently, the method can not recover the H$_2$ column density distribution over the whole column density range. Can the method still be a good indicator of column density over a limited range?

\begin{figure*}
\begin{center}
\includegraphics[trim=0 160 0 0,scale=0.6]{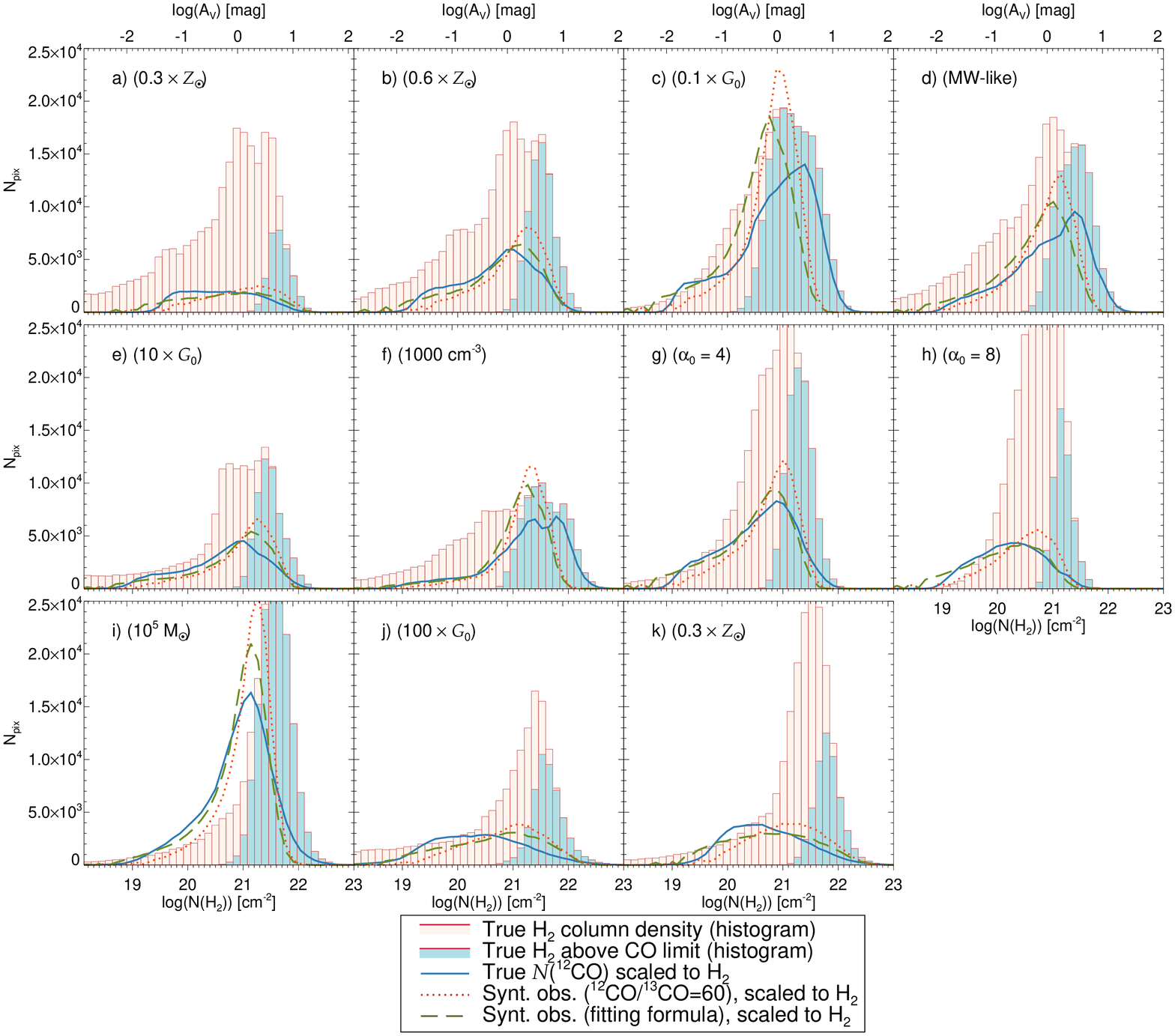}
\end{center}
\caption{H$_2$ column density distributions, taken directly from the simulation (red histogram). The H$_2$ content above the CO detection limit is shown as blue histogram. The blue solid line shows the true $^{12}$CO column density, scaled by a constant abundance ratio to infer H$_2$ mass. The remaining curves represent ``observed'' $^{12}$CO column density distributions, derived under various assumptions (see text in section~\ref{sec:discol}). Note that the $T_{\textrm{b}}^{12} \geq 0.6$ K and $T_{\textrm{b}}^{13} \geq 0.3$ K detection limit corresponds approximately to an inferred H$_{2}$ column density of $4 \times 10^{18}\,{\textrm{cm}^{-2}}$ and a true H$_{2}$ column density of around $10^{21} \, {\textrm{cm}^{-2}}$.}
\label{fig:coldist}
\end{figure*}

To answer this question, we compare the true H$_{2}$ column density distribution to those inferred using the \emph{$\textrm{RD2010}_{\textrm{col}}$} method. We also calculate the true $^{12}$CO column density distribution from the simulation. Fig.~\ref{fig:coldist} shows the different column density distribution for all eleven models.
The red histogram represents the true H$_{2}$ distributions, while the blue histogram shows the H$_2$ column density distribution of regions, where both the $^{12}$CO and $^{13}$CO emission are above the detection threshold. In each case these follow a relatively narrow distribution, peaking between $10^{21}$ and $10^{22} \textrm{cm}^{-2}$.
The $^{12}$CO column densities are transformed to H$_{2}$ column density using a constant $^{12}$CO/H$_2$ ratio. We assume that all carbon atoms are incorporated in CO molecules, thus the abundance ratio is $2 \times x_{\textrm{C,tot}} = 2.8 \times 10^{-4}$ in the case of solar metallicity and it scales linearly with the adopted metallicity. The factor of two is due to the definition of the H$_2$ fractional abundance, which is $x_{\textrm{H}_{2}} = 0.5$ for fully molecular gas.
The blue solid line indicates the distribution of H$_{2}$ column densities we obtain if we apply this procedure to the $^{12}$CO column density distribution taken directly from the simulation.
The remaining curves represent the two distributions inferred from the CO isotope emission maps. In one case, we assume a constant $^{12}\textrm{CO}/^{13}\textrm{CO}$ isotopic ratio equal to 60 (red dotted line), while in the other, a $^{13}$CO column density dependent isotopic ratio (green long-dashed line) is adopted. The former choice is consistent with the frequently assumed value in observations. The latter is a result from numerical simulations and is given in Section~3.4 and Table~3 in \citet{Szucs2014}.

The most striking feature of Fig.~\ref{fig:coldist} is that the CO-inferred distributions fall short of the true H$_{2}$ distribution over the whole column density range. Their PDF peak locations are shifted, on average, to an order of magnitude lower column densities and their distributions are a factor of 2-3 wider than those of the true H$_{2}$ column density above the limit.

The true $N$($^{12}$CO) distribution -- scaled with the given $^{12}$CO/H$_{2}$ abundance ratio -- (blue solid line) only follows the true H$_{2}$ distribution in the high column density wing. This indicates that the chosen $^{12}$CO/H$_2$ ratio is not uniform over the whole detectable cloud. In fact, the true abundance value decreases significantly in the low column density lines of sight, thus widening the inferred column density distributions towards lower values. In the case of our fiducial Milky Way-like model (run d), the mean $^{12}$CO fractional abundance ($x_{^{12}\textrm{CO}}$, the number density ratio with the hydrogen nuclei per unit volume; note that this is half of the $^{12}$CO/H$_2$ ratio when the hydrogen is fully molecular) changes between $1.4 \times 10^{-4}$ (i.e. all carbon in CO) at $A_{\textrm{V}}$ values greater than 2 mag, to a $\textrm{few} \times 10^{-5}$ at visual extinctions between 1 and 2 mag. The main reason for the abundance gradient is the gradual build up of CO shielding from the dissociative ISRF. At lower column densities, where photodissociation is efficient, C$^{+}$ dominates the carbon budget and the remaining small amount of CO follows a different correlation with H$_{2}$ than at high column densities.
Note also, that in reality the true CO distribution might not trace the H$_{2}$ column density distribution even at the highest column densities. High column density sight-lines are often associated with high volume densities, where freeze-out to dust grains depletes the gas-phase CO (a process not considered in the present simulations). For observationally motivated discussions of the cloud depth-dependent CO abundance we refer to the studies of \citet{Pineda2008}, \citet{Lee2014} and \citet{Ripple2013}.

In addition to the error inherited from the incorrect guess for the $^{12}$CO/H$_{2}$ abundance ratio, the H$_2$ column densities derived from the CO emission (red dotted and green long-dashed lines) are also affected by the insufficient detection sensitivity at the low column density end, and saturated, optically thick $^{13}$CO emission at the high column density end.

The former is marked by a cutoff in the inferred\footnote{Note that the true H$_{2}$ column density for these sight lines is much higher, $\sim 10^{21} \: {\rm cm^{-2}}$; the difference between the inferred and real values arises because the mean CO abundance along these sight lines tends to be much smaller than $10^{-4}$.}  H$_{2}$ column density around $N_{\rm H_{2}} = 4 \times 10^{18} \: {\rm cm^{-2}}$  and fluctuations in the PDF at slightly higher column densities, while the latter is shown by the cut-off at the high end. In the cases of strong radiation fields or low metallicities (e.g. runs a, e, j, k), where much of the low density CO is destroyed, the \emph{$\textrm{RD2010}_{\textrm{col}}$} method provides a good representation of the high column density CO distribution.
It is also clear from Fig.~\ref{fig:coldist}, that the adopted $^{12}$CO/$^{13}$CO isotope ratio becomes important in the intermediate column density range ($10^{19} \textrm{cm}^{-2} < N(\textrm{H}_{2}) < 10^{21} \textrm{cm}^{-2}$, equivalent to $4 \times 10^{14} \textrm{cm}^{-2} < N(^{12}\textrm{CO}) < 4 \times 10^{16} \textrm{cm}^{-2}$), where the observations can trace the true CO column density (although not the H$_{2}$ column density) well.

We conclude that the \emph{$\textrm{W2009}_{\textrm{col}}$} and \emph{$\textrm{RD2010}_{\textrm{col}}$} methods are strongly affected by abundance variations at lower column densities and to a lesser extent  by high optical depths at high column densities. Therefore, the methods provide poor estimates for the molecular mass. Nevertheless, they reproduce the CO column densities relatively well over a limited column density range \citep[e.g.][]{KennicuttEvans2012}, if corrections for the depth dependent $^{12}$CO/$^{13}$CO isotope ratio are applied \citep[see Sections 3.3 and 3.4 in][]{Szucs2014}.
In general, however, the H$_{2}$ column density inferred from CO emission does not follow the true distribution. The strong small-scale variations of the CO abundance account for the dominant source of error.

We also point out, that even when dust emission is used to infer the H$_2$ column density distribution, i.e. the abundance variations and the optical depth do not play significant roles, the recovery of the true distribution might be hindered by detection noise, line of sight contamination, field selection and incomplete sampling in interferometric measurements \citep{Ossenkopf2016}.

\subsection{Why does the virial mass estimate work?} \label{sec:disvir}

We show that the virial mass estimates are reliable measures of the H$_{2}$ mass in most simulations, but why is this technique so robust within the covered parameter range, even when the cloud is out of virial equilibrium?

The virial mass analysis relies on 3 main requirements. First, the CO line width should trace the overall velocity dispersion. Second, the cloud should be close to virial equilibrium, and so the velocity dispersion should be proportional to the cloud mass. Finally, the radial mass distribution profile should follow a power law. The exponent of the power law can not be readily measured. In our case, as often in observational studies, the power law exponent is assumed to be $\gamma = 1$.

\begin{figure*}
   \centering
   \begin{subfigure}[b]{0.4\textwidth}
       \includegraphics[trim=45 20 0 30, scale=0.46]{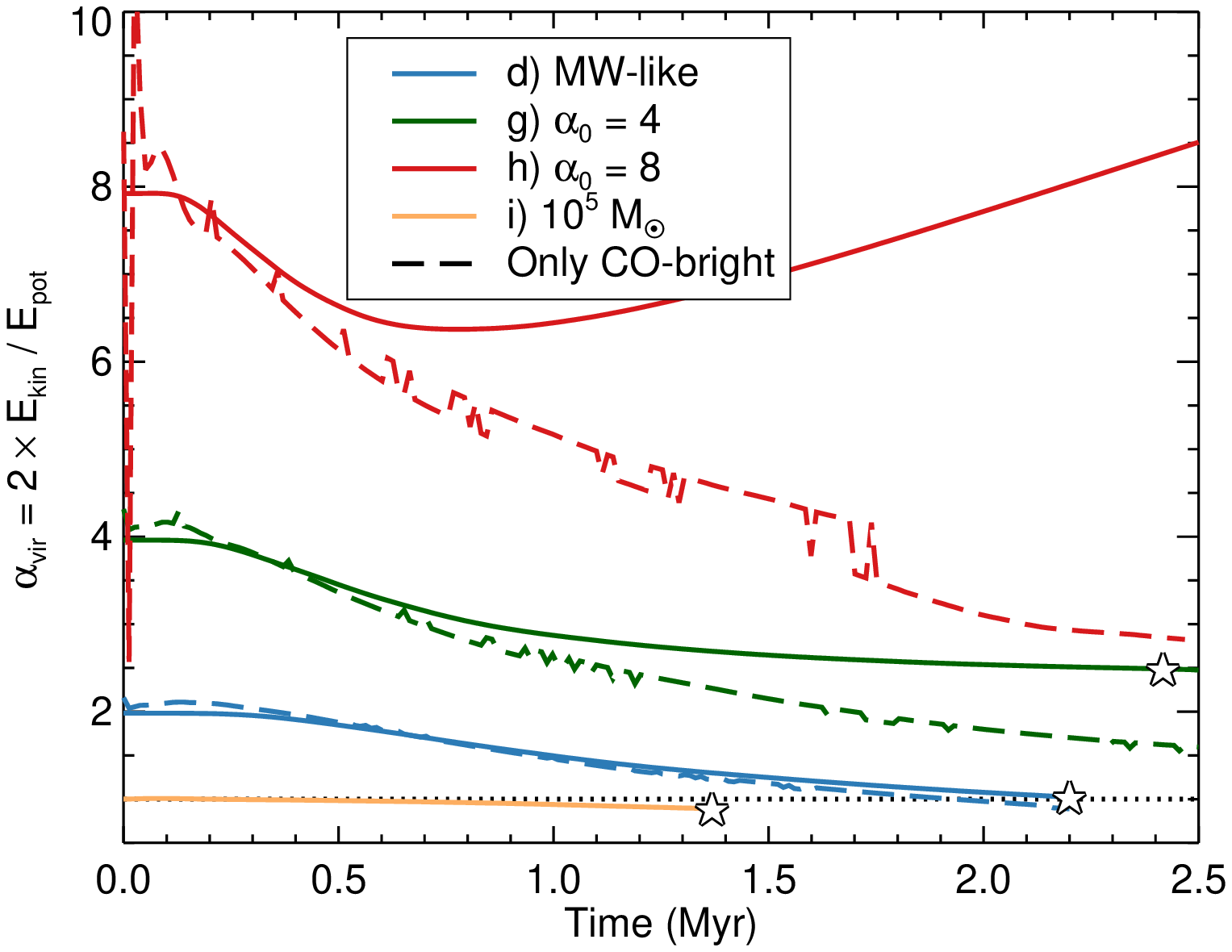}
                \label{fig:alpha}
   \end{subfigure}
~
   \begin{subfigure}[b]{0.4\textwidth}
       \includegraphics[trim=30 20 0 30, scale=0.46]{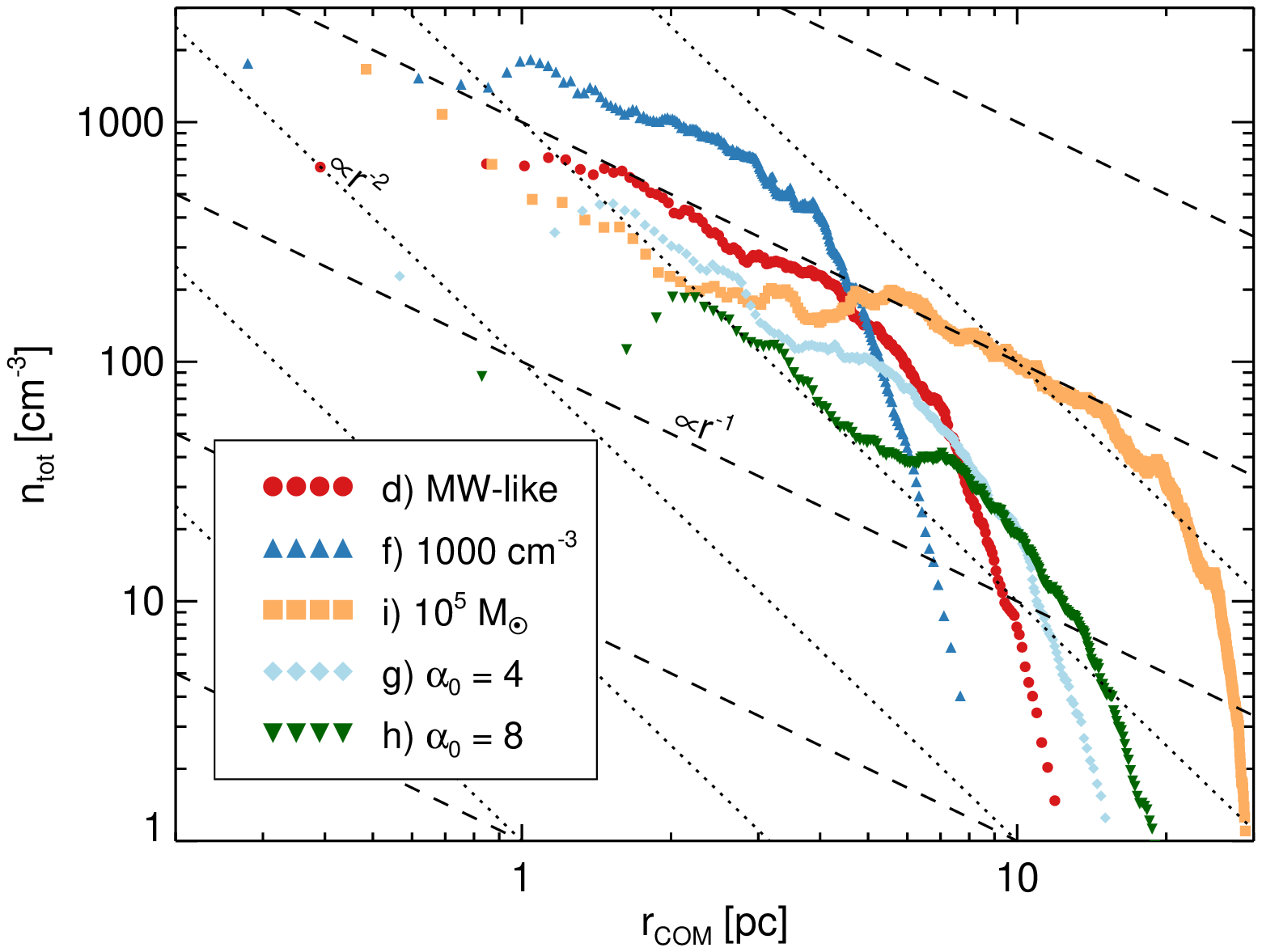}
                \label{fig:radprof}
   \end{subfigure}
\caption{Left panel: the virial parameter defined according to Equation~\ref{eq:vir} as a function of time for the high mass and low mass clouds. The dotted line indicates the equilibrium virial parameter value. The dashed lines show the virial parameter of the CO-bright gas (where $x_{\textrm{CO}} \geq 10^{-5}$), the colour indicates the corresponding model.
We emphasise that the gradual decrease of the virial parameter is expected, because we do not include any sources of turbulent driving in the simulations. See the main text for the explanation of the increase in case of simulation (h).
Right panel: the radial number density profile of the modelled clouds compared to the theoretical assumption of power law indices often used in virial mass measurements. r$_{\textsc{com}}$ denotes the distance from the centre of mass of the cloud.}
\label{fig:virial}
\end{figure*}

Table~\ref{tab:linewidth} compares the mass weighted, true velocity dispersion ($\sigma_{\textrm{1D,intrinsic}}$\footnote{$\sigma_{\textrm{1D,intrinsic}}^2 = \sum v_z(i,j,k)^2 \rho(i,j,k) dV / \sum \rho(i,j,k) dV$, where $v_z$ is the velocity component along the sight line. The summation is for the $i$, $j$, $k$ indexes of the 3 dimensional grid, that is used for the radiative transfer calculation. $\rho$ denotes the density of a given cell and $dV$ is the cell volume. The units are in CGS.}), of the whole cloud, measured along the line of sight, to those inferred from the $^{12}$CO and $^{13}$CO line widths. For both observational indicators, the line-width inferred velocity dispersion is consistent with the true value within $\sim30$-$40$ per cent. Similarly to \citet{MacLaren1988}, we find that the $^{13}$CO velocity dispersion provide a better estimate for the simulated (mass-weighted) line of sight velocity dispersion. In case of the $10^4\,M_{\odot}$ cloud models, the $\sigma_{\textrm{1D,measured}}$ calculated from both the $^{12}$CO and $^{13}$CO lines overestimates the true value. The percentage error in the line width derived from the $^{13}$CO line is generally a factor of a few smaller than the percentage error in the line width derived from the $^{12}$CO line. For example, for model (d), the $^{12}$CO measurement overestimates the true 1D velocity dispersion by 37.4\%, while the $^{13}$CO measurement overestimates it by only 11.4\%.
For the $10^5\,{M}_{\odot}$ cloud models the inferred line widths systematically underestimate the true values. The error is higher for the $^{13}$CO emission, since it does not trace the bulk of the cloud so well as $^{12}$CO (see Fig.~\ref{fig:maps}).

The simulations cover a range of initial virial parameters (defined according to Equation~\ref{eq:vir}). The temporal evolution of $\alpha_{\textrm{vir}}$ is shown in the left panel of Fig.~\ref{fig:virial}.
The total kinetic energy of the cloud is calculated as the sum of the kinetic energies of individual SPH particles (i.e. $E_{\textrm{kin}} = \sum m v_i^{2}/2$, the particle mass, $m$, is 0.005 $M_{\odot}$ and $v_i$ is the absolute value of the velocity vector of the $i$th particle). The total potential energy is calculated directly in the hydrodynamic simulation to take the irregular mass distribution fully into account.
If $\alpha_{\textrm{vir}}$ is close to 1, then the virial cloud mass estimate is expected to be reliable.
The virial parameter of the $10^{4} M_{\odot}$ simulations with initial $\alpha_{0} = 2$ (blue solid line) stagnates during the first $\sim0.3$ Myr, then falls off as a power law. The cloud is roughly in virial equilibrium when sink particles first begin to form. In the case of the more massive cloud with initial $\alpha_{0} = 1$ (yellow solid line), $\alpha_{\textrm{vir}}$ decreases below the equilibrium value slowly, on a Myr time scale. The $\alpha_0 = 4$ model behaves similarly to the fiducial model. A qualitative change in behaviour is evident for the $\alpha_0 = 8$ model. The virial parameter increases after an initial decrease in the first 0.5 Myr. This due to the fact that the overall gravitational potential energy decreases with a higher rate (as the cloud is dispersed) than the kinetic energy decays.

The virial parameters discussed above are calculated for the whole cloud. The (synthetic) observations, however, trace only the regions, which are bright in CO emission. Therefore, the CO emission might not give us information on the virial state of the whole cloud.
The dashed curves on the left panel of Fig.~\ref{fig:virial} show the virial parameter of the CO-bright segment of the clouds (see also Table~\ref{tab:viralpha}). We consider SPH particles as being ``CO-bright'', if their $^{12}$CO abundance is higher than $10^{-5}$ (relative to H nuclei number density). In reality, the brightness depends also on the local gas temperature and on the excitation properties. Due to the complexity of mapping between a regular grid, on which the emissivity is calculated, and the unstructured assembly of SPH particles, used for the kinetic and potential energy measurement, we rely on the simplification.
The CO-bright- and cloud-wide virial parameters of the Milky Way-like model (d) evolve similarly. The similarity remains for the $\alpha_{0} = 4$ model, but the $\alpha_{\textrm{vir}}$ of the CO-bright gas decreases more rapidly and at the analysed evolutionary time it is close to 2. The CO-bright virial parameter of the $\alpha_0 = 8$ model, in contrast to the virial parameter of the whole cloud, decreases rapidly with time. At the time of the analysis it is about 3. The trend is comparable to the differences of virial mass estimates, obtained for the models. This is explained by the enhanced CO photodissociation in dispersing clouds with initially large virial parameter. CO is only retained in self-gravitating dense clumps, where the kinetic energy quickly dissipates.
The fact that the virial parameter of CO-bright regions is within a factor of few of the equilibrium value explains why the virial mass estimate seems to work well for all of these clouds.

The radial density profile of the molecular cloud, $\rho(r)$ or $n(r)$, depending whether the mass or number density is concerned, is the quantity determining the numerical value of the multiplier in equation~\ref{eq:mvir2}. The right panel of Fig.~\ref{fig:virial} shows the average number density of selected clouds as a function of cloud radius. The figure was constructed by dividing the total number of protons in small $\delta r$ thickness spherical shells around the centre of gravity with the shell volume. We compare five representative models: the radial density profile of model (d) is very similar to that of (a), (b), (c), (d) and (e). Similarly, simulation (i) is a good representation of simulations (j) and (h). The initially denser cloud (model f) and the two enhanced virial parameter models (g and h) are plotted individually for comparison.

The $10^{4} M_{\odot}$ models yield relatively flat distributions in the inner 1 pc, then they follow $n(r)\propto r^{-1}$, until the sharp cut-off at the cloud edges. On relatively large size scales, where the bulk of the mass is located, the radial number density profile follows the correlation, assumed in the \emph{$\textrm{RL2006}_{\textrm{vir}}$} and \emph{$\textrm{ML1988}_{\textrm{vir}}$} methods, remarkably well. This is especially curious, because the simulations start with homogeneous and isotropic density distributions, with a mean number density of 300 $\textrm{\textrm{cm}}^{-3}$ (1000 $\textrm{\textrm{cm}}^{-3}$ in case of f).
The $10^{5} M_{\odot}$ runs are very similar, with the exception of a steeper profile in the inner cloud.
We believe that the radial density profile is not an artefact in the simulation, since it is very similar in both the low mass and the high mass runs. For these runs we use different seed fields to produce the initial, random velocity fluctuations and also choose different amplitudes for the turbulent velocities (by changing the cloud mass).

The radial density profile, however, changes with the virial parameter of the cloud: when $\alpha_0 = 2$ the exponent is about -1, while when $\alpha_0 = 8$, the exponent is approximately -2. In the latter case the coefficient in equation~\ref{eq:mvir2} needs to be adjusted to the appropriate value of $698\,\textrm{M}_{\odot}\,\textrm{pc}^{-1}\,\textrm{km}^{-2}\,\textrm{s}^{2}$ \citep{MacLaren1988}. Thus it yields a 67 per cent smaller virial mass estimates. When this correction factor in applied to the virial mass estimates from model (h), we find a better agreement with the CO-bright molecular mass.

In short, the virial mass estimate from the simulations works well because the emission lines do a good job in tracing the true mass weighted velocity dispersion. The virial parameter converges to unity by the epoch of sink particle formation in the $\alpha_0 = 2$ simulations. When the initial virial parameter is high (g and h), the overall $\alpha_{\textrm{vir}}$ value of the cloud is significantly higher than that of the CO-bright regions. The later is about a factor of 2-3 above the equilibrium value. We also find an increasing slope of the radial density profile with the increasing virial parameter. Due to these issues, the \emph{$\textrm{RL2006}_{\textrm{vir}}$} and \emph{$\textrm{ML1988}_{\textrm{vir}}$} methods overestimate the CO-bright cloud mass when the initial virial parameter is much higher than unity. A fraction of the real CO-rich molecular clouds are expected to behave similarly to models (g) and (h), while clouds that are formed in large scale colliding flows might inherit large velocity dispersion, and thus large virial parameters \citep[e.g.][]{Vezquez2007,Kauffmann2013}. CO emission might not be a good indicator for the overall virial state of such turbulent clouds, as it traces only the CO abundant gas that coincides (due to the necessary shielding from UV radiation) with the gravitationally collapsing regions \citep{GloverClark2012a,Clark2012b}. The collapsing regions typically have a virial parameter close to the equilibrium value \citep{Ballesteros-Paredes2006}.

Note also that a fundamental assumption underlying the virial method is that the observed line widths are dominated by turbulent broadening rather than thermal broadening. This is a good approximation for all of the clouds considered in our study, but can potentially break down if we consider clouds in extreme environments such as the Galactic Centre (see Bertram et al., 2016, in prep.).

\begin{table*}
\caption{Intrinsic and measured velocity dispersion. For the details of the calculation we refer to Section 4.1 in \citet{Szucs2014}.}
\label{tab:linewidth}
\begin{center}
\begin{tabular}{ccccccccccc}
  \hline
  \multicolumn{2}{c}{Model} & \multicolumn{2}{c}{$^{12}$CO ($J=1\rightarrow0$)} & \multicolumn{2}{c}{$^{13}$CO ($J=1\rightarrow0$)} \\
        & $\sigma_{\textrm{1D,intrinsic}}$ & $\sigma_{\textrm{1D,measured}}$ & relative deviation & $\sigma_{\textrm{1D,measured}}$ & relative deviation \\
        & ($\textrm{km s}^{-1}$) &  ($\textrm{km s}^{-1}$) & $\%$ & ($\textrm{km s}^{-1}$) & $\%$ \\

  \hline
      a & 0.94  & 0.98 & +4.6  & 0.93 & -0.01 \\
      b & 0.92  & 1.13 & +22.6 & 0.99 & +7.5  \\
      c & 0.87  & 1.37 & +57.2 & 0.97 & +10.9 \\
      d & 0.92  & 1.27 & +37.4 & 1.03 & +11.4 \\
      e & 0.97  & 1.10 & +13.9 & 1.02 & +5.2  \\
      f & 1.09  & 1.50 & +37.9 & 1.25 & +14.8 \\
      g & 0.82  & 1.10 & +34.1 & 0.86 & +5.3  \\
      h & 0.87  & 0.97 & +10.9 & 0.79 & -9.0  \\
      i & 1.71  & 1.67 & -2.4  & 1.29 & -24.6 \\
      j & 1.86  & 1.57 & -15.8 & 1.52 & -18.4 \\
      k & 1.81  & 1.51 & -16.4 & 1.45 & -19.6 \\
 \hline

\end{tabular}
\end{center}
\end{table*}

\subsection{ \texorpdfstring{$A_{\textrm{V}}$-dependent $X_{\textrm{CO}}$ and small-scale variations}
                   {The $A_{\textrm{V}}$-dependent CO-to-H2 factor and its small-scale variations} }

\subsubsection { \texorpdfstring{Recovery of the CO-bright and total H$_2$ masses}
                   {Recovery of the CO-bright H$_2$ mass} }

The {\it GML2001$_{XCO}$} method overestimates the H$_{2}$ mass in runs (a) and (k) (low metallicity) and (g) and (h) (high virial parameter) by more than a factor of a few. This major overestimation is explained by the absence of self-gravity in the simulations of \citet{GloverMacLow2011}. In the absence of self-gravity, it becomes extremely difficult to form regions with local visual extinctions high enough to shield CO against photodissociation by the ISRF when the cloud-averaged extinction is small. Therefore, simulations of low mean extinction clouds (e.g. low metallicity clouds or clouds that are dispersed by high levels of turbulence) that are carried out without self-gravity produce only a small amount of CO, and the resulting CO-to-H$_{2}$ conversion factor is large. On the other hand, models that account for self-gravity can capture the formation of dense, self-gravitating clumps. As they collapse, these clumps increase their local density and extinction, and become important sites of CO formation. Previous simulations of low metallicity clouds have shown that these dense clumps quickly come to dominate the total CO emission \citep{GloverClark2012c,GloverClark2016}. As a result, once these clumps form, the total CO emission becomes much larger than the value predicted by the \citet{GloverMacLow2011} simulations, and consequently the appropriate $X_{\rm CO}$ factor is smaller than the \citet{GloverMacLow2011} prediction.

For the low metallicity models, the \emph{W2010$_{XCO}$} method traces the total H$_{2}$ mass rather than the H$_{2}$ above the CO detection limit. This is not surprising, since the method was calibrated to recover the total H$_{2}$ mass. In any other cases (i.e. when the metallicity is solar) the method returns the galactic $X_{\textrm{CO}}$ factor, independent of the cloud average visual extinction.

Surprisingly, the Galactic $X_{\textrm{CO}}$ factor provides a reasonable estimate of the H$_{2}$ mass above the CO detection limit, even when the metallicity is low (estimate within a factor of three). This might be explained by invoking the argument of \citet{Liszt2010} for close to Galactic $X_{\textrm{CO}}$ factors in diffuse clouds: They separate the $X_{\textrm{CO}}$ factor into two coupled and competing contributions:
\begin{equation}
\frac{1}{X_{\textrm{CO}}} = \frac{W(\textrm{CO})}{N(\textrm{CO})} \times \frac{N(\textrm{CO})}{N(H_{2})},
\end{equation}
where the first factor accounts for radiative transfer and CO molecule excitation effects, while the second accounts for the chemistry. On one hand the $N(\textrm{CO}) / N(\textrm{H}_{2})$ ratio decrease with decreasing metallicity (i.e. less dust shielding leads to more CO photodissociation and even in well-shielded regions there is less carbon and oxygen available); on the other hand, the $W(\textrm{CO})/N(\textrm{CO})$ ratio increases due to the higher gas and excitation temperatures. It can be shown, that the decrease of one factor might be compensated by the other \citep{Pety2008}.
When models (a) and (d) are compared, in fact we find that along the high column density lines of sight, where the CO abundance changes by a factor of 3, the line of sight mass-weighted gas temperature changed with a similar factor.

\begin{figure*}
\begin{center}
\includegraphics[trim=0 250 0 0,scale=0.65]{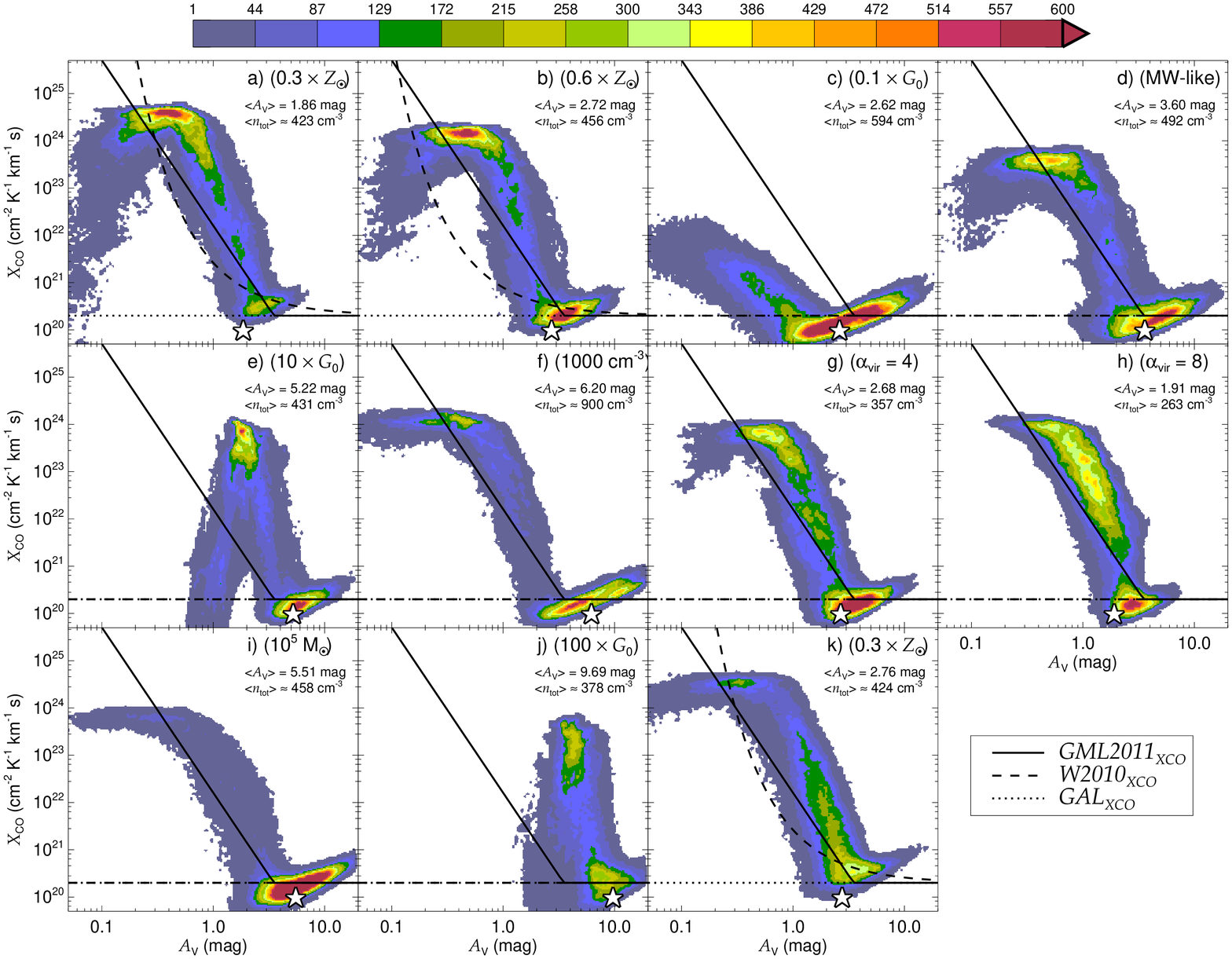}
\end{center}
\caption{The true $X_{\textrm{CO}}$ factor in the simulations as a function of visual extinction, for the pixels where $W$(CO) $>$ 0 K km s$^{-1}$. The figure shows the two-dimensional histogram of pixels which have a given $A_{\textrm{V}}$ -- $X_{\textrm{CO}}$ combination (see colour bar). The star symbol indicates the cloud average visual extinction. The predicted cloud average $A_{\textrm{V}}$ -- $X_{\textrm{CO}}$ factor correlations of the \emph{GML2011$_{XCO}$}, \emph{W2010$_{XCO}$} and \emph{GAL$_{XCO}$} are over-plotted. These are evaluated at cloud average visual extinction ($\langle {A}_{\textrm{V}} \rangle$; marked by the star symbol). The $\langle {A}_{\textrm{V}} \rangle$ is the average visual extinction of pixels that are above the CO detection threshold (see section~\ref{sec:detlim}).}
\label{fig:xcoall}
\end{figure*}

Another issue that needs to be mentioned is the combined effect of increasing CO-dark H$_2$ mass fraction with decreasing metallicity (and/or increasing ISRF strength) and the application of a brightness temperature detection limits to the CO emission. Due to the increasing CO-dark mass fraction $X_{\textrm{CO}}$ also increases. By imposing a detection limit, we discard these CO-dark (or rather CO-faint) lines of sight and hence weaken the metallicity (or ISRF) dependence of the $X_{\textrm{CO}}$ factor \citep[see e.g. the discussion in][]{Bolatto2013}. As a result, the observed $X_{\textrm{CO}}$ factor remains close to the galactic value.

\subsubsection { \texorpdfstring{$X_{\textrm{CO}}$ factor on sub-parsec scales}
                   {$X_{\textrm{CO}}$ factor on sub-parsec scales} }

It is widely accepted in the literature \citep[e.g.][]{GloverMacLow2011,Shetty2011a,Shetty2011b,Bolatto2013} that the $X_{\textrm{CO}}$ conversion factor breaks down on small scales. The clumpy nature of molecular clouds leads to an inhomogeneous radiation field within them. This in turn regulates the CO abundances and excitation conditions, resulting in large changes on the sub-parsec scales. None of the $X_{\textrm{CO}}$ methods discussed in this paper claim to provide reliable conversion factors for individual pixels. In fact, they are recommended to be used on the cloud-averaged quantities, in order to smooth out any small scale fluctuations. Arguably, the $X_{\textrm{CO}}$ factor should not even be used on the scale of individual clouds, but instead only as averages over whole populations \citep{KennicuttEvans2012}. The behaviour of the cloud average $X_{\textrm{CO}}$ factor, however, must reflect systematic changes on (sub-)parsec size scales. To investigate this, we plot the true $X_{\textrm{CO}}$ factor (defined as the pixel-wise ratio of the true H$_2$ column density and the integrated $^{12}$CO emission, where $W(^{12}\textrm{CO}) > 0$ K km s$^{-1}$) against the visual extinction, measured along the corresponding sight-lines (Fig.~\ref{fig:xcoall}). The colour indicates the probability of finding a sight-line (on the pixel scale) with the given parameter combination. The figure also shows the adopted $X_{\textrm{CO}}$ factors (lines) and the mean visual extinction of the cloud above the CO detection limit (star).

\begin{figure}
\begin{center}
\includegraphics[trim=0 30 0 0,scale=0.45]{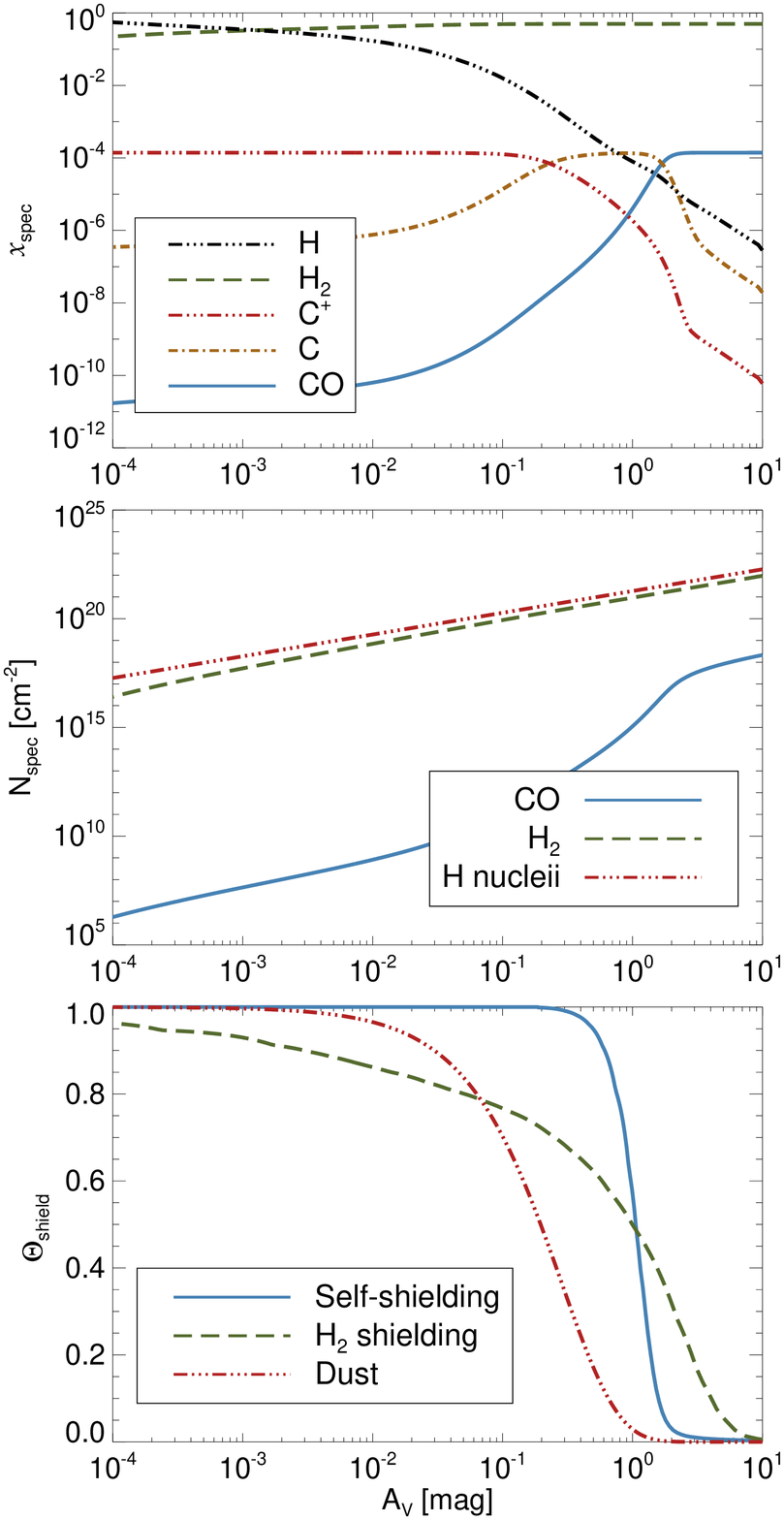}
\end{center}
\caption{Simple model to demonstrate the importance of CO shielding mechanisms as function of visual extinction. The density distribution is adopted according to the Bonnor-Ebert sphere solution. The depth dependent abundances and shielding factors are calculated iteratively.}
\label{fig:shield}
\end{figure}

Fig.~\ref{fig:xcoall} shows a qualitatively similar visual extinction dependence for the conversion factor in all models. This behaviour is partly driven by the changing dominant CO and H$_2$ shielding processes with the increasing column density. The exact column density dependence of the efficiency of CO shielding processes, however, depend on the 3D density distribution and chemical composition of the cloud. Fig.~\ref{fig:shield} illustrates this dependence in case of a 1 dimensional, static Bonnor-Ebert sphere model, with linked and iteratively solved shielding and chemical models. The adopted Bonnor-Ebert sphere has a central density of $3 \times 10^{6}\,\textrm{cm}^{-3}$. The abundances and column densities of species are found iteratively; first the column density is calculated according to a constant initial abundance. Then with the shielding, determined by the column density, the chemical model is run. The new column densities are determined using the new abundances, and so on. The shielding coefficients ($\Theta_{\textrm{CO}}, \Theta_{\textrm{H}_2}, \Theta_{\textrm{dust}}$: CO self-shielding, shielding by H$_2$ lines and dust shielding, respectively) set the CO photodissociation rate, $R_{\textrm{CO,thick}}$ according: $R_{\textrm{CO,thick}} = \Theta_{\textrm{CO}} \Theta_{\textrm{H}_2} \Theta_{\textrm{dust}} R_{\textrm{CO,thin}}$, where $R_{\textrm{CO,thin}}$ is the unattenuated CO photodissociation rate \citep[see e.g.][]{Visser2009b}. For the references of the shielding coefficients see Section~\ref{sec:sims}.
Considering the depth-dependent shielding factors in this simplified model, we distinguish 4 characteristic ranges in the $A_{\textrm{V}}$--$X_{\textrm{CO}}$ distributions (in Fig.\ref{fig:xcoall}):

\renewcommand{\labelenumi}{(\arabic{enumi})}

\begin{enumerate}

\item Very low visual extinction ($A_{\textrm{V}} < 0.2\,\textrm{mag}$): The $X_{\textrm{CO}}$ factor increases due to the quickly increasing H$_{2}$ column density and the very low ($x_{\textrm{CO}} \approx 10^{-11}$) and approximately unchanged CO abundance. CO molecules are weakly shielded from the interstellar UV radiation and the dominant shielding mechanism is due to the overlapping Werner-band H$_{2}$ absorption lines.

\item Low visual extinction ($0.5\,\textrm{mag} < A_{\textrm{V}} < 1\,\textrm{mag}$):
The factor levels off and stays nearly constant over a narrow $A_{\textrm{V}}$ range. The H$_{2}$ and the CO column densities are increasing similarly. In this range the dominant shielding mechanism is the dust shielding for both molecules.

\item Translucent region (around 1-2 mag): the carbon is converted in a relatively narrow extinction range from ionised and atomic forms to CO molecules. This leads to a rapid increase in the efficiency of CO self-shielding, and thus a positive feedback to the CO conversion. The CO emission also increases rapidly, leading to a decreasing $X_{\textrm{CO}}$ factor.

\item High visual extinction ($A_{\textrm{V}} >$ few mag):
      The $X_{\textrm{CO}}$ factor increases again due to the increasing H$_{2}$ column density but saturating (optically thick) CO emission. The change between the negative and positive slops of the conversion factor takes place at a few $A_{\textrm{V}}$ visual extinction and an $X_{\textrm{CO}}$ close to the Galactic value. In typical molecular clouds (i.e. where sufficient shielding is available), a large fraction of the gas mass is located at this transition, thus a single $X_{\textrm{CO}}$ value might be used to derive cloud masses.

\end{enumerate}

Generally, the distributions are roughly bimodal, with most pixels falling in ranges (2) and (4). In range (2) the typical $X_{\textrm{CO}}$ factors are between a few times $10^{23}$ and $10^{24}$ $\textrm{cm}^{-2}\,\textrm{K}^{-1}\,\textrm{s}$, depending on the metallicity (see panels a, b and d). In range (4) the conversion factor is close to the galactic $X_{\textrm{CO}}$ factor.
The exact transitional visual extinction value and to some extent the shape of the distribution depend on the physical conditions. With decreasing metallicity, the distribution shifts towards higher $X_{\textrm{CO}}$ values and more pixels fall into region (2). The increasing radiation field strength leads to more efficient CO destruction and a distribution shifted towards higher $A_{\textrm{V}}$ values. The galactic $X_{\textrm{CO}}$ factor provides a good measure of the H$_2$ mass of the cloud regions, where the dominant CO shielding mechanism is the self-shielding. The saturation of CO emission leads to only small deviations from this value.

\section{Summary} \label{sec:sum}

In this paper we investigate the validity of the most frequently used methods to determine cloud mass and column density based on CO emission measurements. We analyse hydrodynamic simulations with realistic thermal balance and chemical modelling and explore a range of metallicities, ISRF strengths, cloud densities, cloud masses and virial parameters. Emission maps are constructed from the simulated (molecular) number densities, velocity fields and gas temperatures, using the LVG approximation. No observational noise is added to the emission maps, but we only consider PPV voxels in the analysis, which are above a given detection limit. In the case of methods that use both the $^{12}$CO and $^{13}$CO emission, only voxels above the respective $0.6\, \textrm{K}$ and $0.3\, \textrm{K}$ detection thresholds are considered. In the case of methods that rely solely on the $^{12}$CO emission, we consider all voxels above $0.6\, \textrm{K}$ brightness temperature. The chosen detection limits are comparable to the $3\,\sigma_{\textrm{rms}}$ levels of current-day single dish observations.

We investigate three main methods: (1) the inference of H$_{2}$ column density and mass from the $^{13}$CO column density measurement, (2) the virial analysis and (3) the CO emission-to-H$_{2}$ column density conversion factor (i.e. $X_{\textrm{CO}}$ factor). We also test slightly different, alternative implementation of these methods. The main results can be summarised as follows:

\renewcommand{\labelenumi}{(\roman{enumi})}

\begin{enumerate}

  \item All methods except the $^{13}$CO column density measurement provide mass estimates within a factor of 2 error over a large range of parameters, as long as the metallicity is not too low. If the conditions are similar to those in the Milky Way (for which the methods were originally calibrated) the agreement is even better.

  \item The LTE column density measurement of $^{13}$CO might trace the true CO column density distribution well within the $4 \times 10^{14}\,\textrm{cm}^{-2} < N(^{12}\textrm{CO}) < 4 \times 10^{16}\,\textrm{cm}^{-2}$, or $10^{19}\,\textrm{cm}^{-2} < N(\textrm{H}_{2}) < 10^{21}\,\textrm{cm}^{-2}$ range, if the fitting formula presented in \citet{Szucs2014} for the $^{12}$CO/$^{13}$CO ratio is used. However, the method is a bad indicator for the true H$_2$ column density and thus the overall cloud mass.  The reason for the poor performance is the fact that the method assumes a common, fixed CO abundance ratio in each pixel of the resolved cloud. In reality, the abundance ratio is not homogeneous and is expected to vary from pixel-to-pixel (or beam-to-beam) \citep[see also][]{Ripple2013}.

  \item The virial mass estimate, determined from CO emission, provides a reliable measure for the CO-bright H$_{2}$ mass of simulated clouds with initially virial parameters close to equilibrium (models a to f and i to k). On one hand, the virial parameter relaxes or remains close to equilibrium by the time sink particle formation begins. On the other hand, the turbulence and self-gravity develop a radial density distribution which is close to the assumed $n(r) \propto r^{-1}$ distribution, and the emission line width traces the true line of sight velocity dispersion relatively well.
In the case of models with high initial virial parameter (g and h), the mass is slightly overestimated. This small excursion, however, does not reflect the large deviation from virial equilibrium. The method yields a reasonable estimate, because the virial parameter of the CO-bright regions of the cloud (which is actually traced by the observations, see Fig.~\ref{fig:virial}) approaches the equilibrium value ($\alpha_{\textrm{vir}} = 1$) more rapidly than the cloud as a whole. Additionally, we find that the radial density profile of the cloud steepens with increasing $\alpha_{\textrm{vir},0}$ (see right panel of Fig.~\ref{fig:virial}). When the coefficient in Equation~\ref{eq:mvir2} is adjusted to the steeper profile, the methods yield a more accurate mass estimate.
We also find that the purely $^{12}$CO emission based \emph{$\textrm{RL2006}_{\textrm{vir}}$} method preforms systematically better than the \emph{$\textrm{ML1988}_{\textrm{vir}}$} method, which uses both the $^{12}$CO and the $^{13}$CO emission.

  \item The observed virial parameter, defined as the ratio of the virial mass and the CO-bright or total cloud mass, are in all cases close to unity (see Table~\ref{tab:viralpha}). In particular, they are always subvirial, when compared to the total mass (i.e. what the dust column density would measure). Consequently, the virial parameter estimates might only provide a lower limit on the true $\alpha_{\textrm{vir}}$ of the cloud. Thus clouds, observed to be (sub-)virial, might in fact be stable to gravitational collapse or even expand on large scales.

  \item The $X_{\textrm{CO}}$ factor provides a good molecular mass estimate over a range of values in ISRF strength, initial density, cloud mass and virial mass, unless the metallicity is low. This might be a result of the excitation temperature balancing the CO abundance \citep{Liszt2010}. And/or it might be a result of a bias, originating from the adoption of a CO brightness temperature limit \citep{Bolatto2013}.
At low metallicities the galactic $X_{\textrm{CO}}$ value consistently underestimates the true H$_{2}$ mass, usually by a factor of 3, while the $\langle A_{\textrm{V}} \rangle$ and metallicity dependent methods of \citet{Wolfire2010} and \citet{GloverMacLow2011} tend to overestimate the true CO-bright mass, occasionally by more than an order of magnitude. We recommend the prescription of \citet{Wolfire2010} for a metallicity-dependent $X_{\textrm{CO}}$ factor, with the caveat that the method indicates the total molecular mass and not only its CO-bright fraction.

  \item We find that the mass fraction of CO-dark gas (i.e. molecular gas not traced by CO) depends on metallicity, radiation field strength, and virial parameter, but also on the detection threshold. The inferred value ranges from 20 to 50 per cent, when the metallicity is low or the ISRF is high or the virial parameter is high, to about 5 per cent, when the ISRF is weak or the mean density of the cloud is high. In case of Milky Way-like conditions ($1 \times G_0$ ISRF, solar metallicity), the CO-dark molecular mass fraction is 12 per cent. The CO-dark fraction is a factor of 1.4-1.5 larger when the emission of both CO isotopes is required to be above the threshold, compared to when only the $^{12}$CO brightness threshold is imposed.
  Due to the fact that we simulate isolated MCs (which initially are spherically symmetric and have homogeneous density) without considering a more extended and diffuse ISM component, we do not expect the numerical values of the dark molecular gas fraction to reflect those of the real interstellar medium \citep[for further discussion see][]{Smith2014}, but the trends should be instructive.

  \item The results show a weak dependence on viewing angle and a moderate cloud-to-cloud variation. When alternative realisations of the physical conditions are examined, we find a less than a factor of two deviation in the derived H$_2$ masses (see Appendix~\ref{appdx:viewangle}).

\end{enumerate}

\section*{Acknowledgements}
We thank the anonymous referee for helping to improve the quality and clarity of the manuscript with a thorough report.
The authors acknowledge support from the Deutsche Forschungsgemeinschaft via SFB project 881 ``The Milky Way System'' (sub-projects B1, B2, B3, B8). R.S.K. and S.C.O.G. also acknowledge support from the Deutsche Forschungsgemeinschaft via SPP 1573, ``Physics of the Interstellar Medium''  (grant number GL 668/2-1).
The numerical simulations were partly performed on the KOLOB cluster at the University of Heidelberg and partly on the Milky Way supercomputer, funded by the DFG through Collaborative Research Center (SFB 881) ``The Milky Way System'' (subproject Z2), hosted and co-funded by the J\"ulich Supercomputing Center (JSC).
L. Sz. acknowledge support from the European Research Council grant PALs (project number 108477).
R.S.K. acknowledges support from the European Research Council under the European Community\'s Seventh Framework Programme (FP7/2007-2013) via the ERC Advanced Grant STARLIGHT (project number 339177).

\bibliographystyle{mn2e}
\bibliography{references}

\appendix

\section{Interpolation to regular grid} \label{appdx:restest}

\begin{figure}
\begin{center}
\includegraphics[trim=0 35 0 0, scale=0.5]{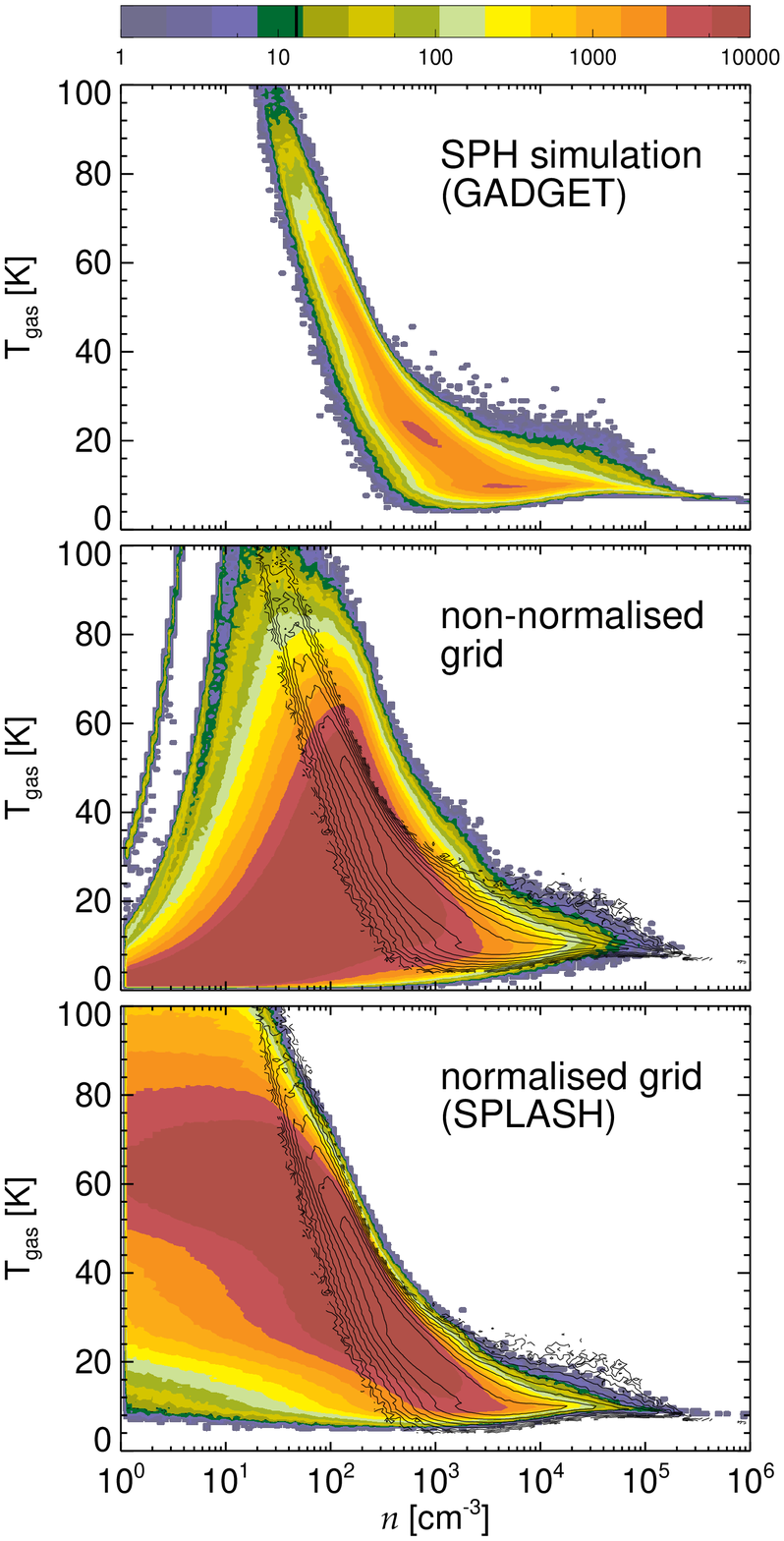}
\end{center}
\caption{Comparison of the density -- gas temperature distribution in the SPH simulation (model d) and on the $512^3$ resolution grid. The contour lines in the middle and lower panels represent the SPH distribution. In this paper we use normalised interpolation. The colour indicates the number of SPH particles or grid elements with a given parameter combination.}
\label{appdxfig:interpol}
\end{figure}

The variable smoothing length SPH formalism of hydrodynamics relates the spatial resolution of the simulation to the local density (i.e. the density of Lagrangian particles). In molecular cloud simulations, where multiple size scales need to be considered (from tens of parsec scales in dilute gas to the dense protostellar cores on the order of thousands of astronomical units in size), this feature is often required.
However, the radiative transfer code applied here, {\sc radmc-3d}, works on gridded input data. By interpolating the adaptive spatial resolution SPH data to a fixed size grid, we unavoidably lose information and introduce systematic errors.

The standard method of SPH interpolation assumes that there are sufficient amount of particles around a point of interest, that:
\begin{equation}
\sum_{j=1}^{N} \frac{m_{j}}{\rho_{j}} W(|r-r_{j}|,h) \approx 1,
\end{equation}
where $N$ is the total SPH particle number, $m_{j}$, $\rho_{j}$ and $r_{j}$ are the mass, density and position of particle $j$ and $h$ is the smoothing length.
This criterion hold by design at the SPH particle locations. In most instances it also applies for arbitrary locations within well resolved, dense regions. At the cloud edge, however, the number of particles within a distance of $h$ from an arbitrary location might be small, yielding a sum much less than unity.

If a non-normalised interpolation scheme is used \citep[e.g.][]{Szucs2014,Glover2015a}, the sum is not corrected, leading to an underestimation of the interpolated quantities (e.g. temperature). In case of normalised interpolation, the quantities are divided by the above sum at the position. This corrects for the overestimation, but introduces artefacts of its own. Individual particles might start to dominate the interpolated quantity. For further explanation and examples we refer to \citet{Price2007}.

\begin{figure}
\begin{center}
\includegraphics[trim=0 35 0 0, scale=0.5]{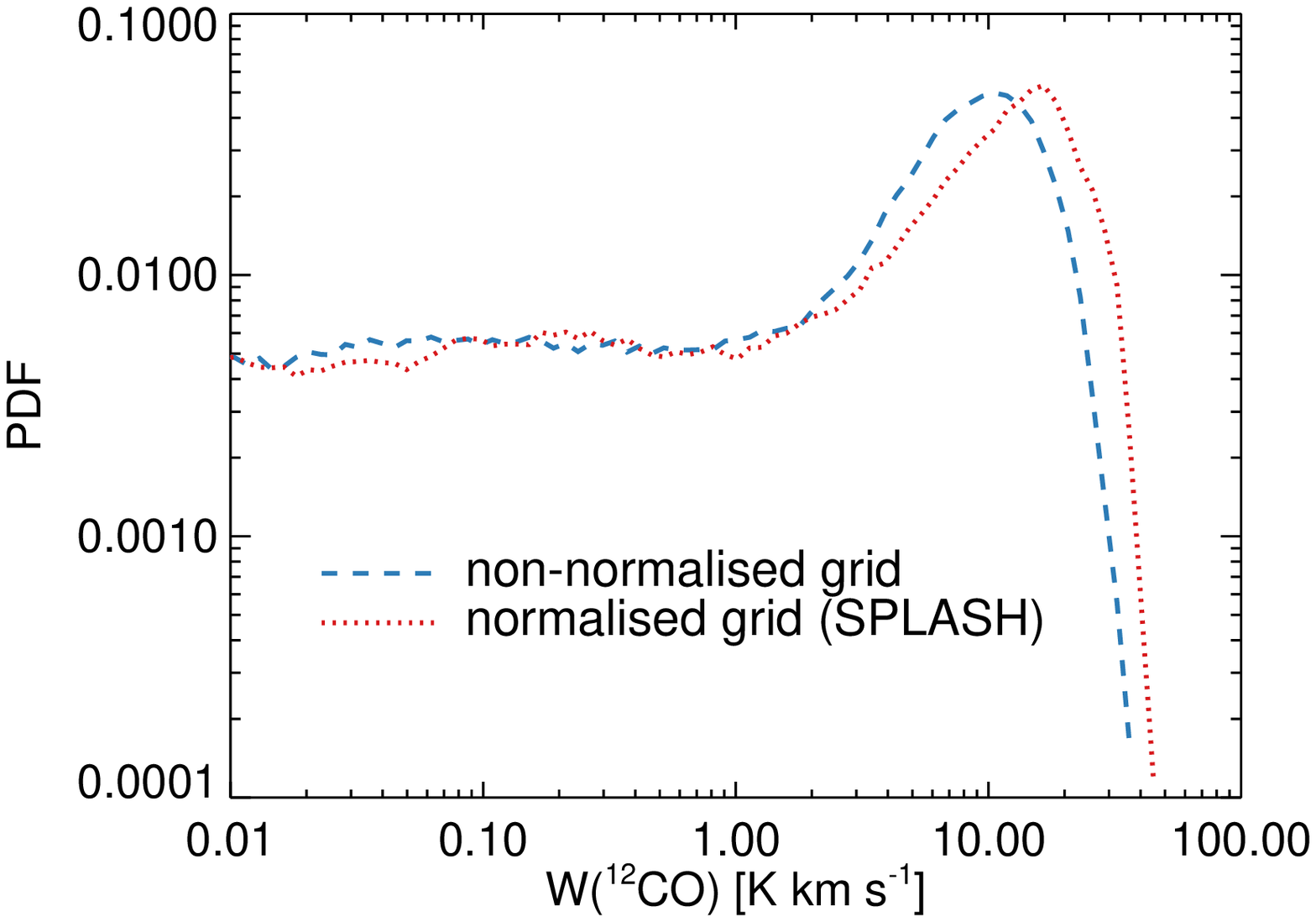}
\end{center}
\caption{$^{12}$CO ($J=1\rightarrow0$) integrated emission probability density distribution of model d) using the two interpolation schemes.}
\label{appdxfig:Wcointerpol}
\end{figure}

We investigate how the choice of the interpolation scheme influences the interpolated quantities and the emission maps.
Fig.~\ref{appdxfig:interpol} illustrates the difference between the approaches in the case of the gas temperature. The upper panel shows the density -- gas temperature distribution of the SPH particles. The middle and bottom panels show the interpolated quantity at the resolution of $512\times512$ pixels and $h=0.032$ pc.
When no normalisation is used the temperature at low density is underestimated and can reach 0 K, even in regions ($n \approx 300 \textrm{cm}^{-3}$) where the emission might otherwise contribute measurably to the total luminosity of the cloud. At the same time, in some grid cells the temperature is overestimated at densities as high as a few thousand particles per cm$^{3}$.

\begin{table}
\caption{Total mass on the SPLASH grid as a function of resolution and the total SPH particle mass within the analysed domain. The comparison is done in case of model d).}
\label{tab:gridmass}
\begin{center}
\begin{tabular}{cc}
  \hline
      Resolution  &  Mass [$\textrm{M}_{\odot}$] \\
  \hline
      $128^3$ & 37000  \\
      $256^3$ & 11015  \\
      $512^3$ & 9351   \\
      $768^3$ & 9292   \\
 \hline
      SPH mass & 9250  \\
 \hline
\end{tabular}
\end{center}
\end{table}

When normalised interpolation is used these artefacts disappear. However, the scheme has its own disadvantage, namely it overestimates the mass. Table~\ref{tab:gridmass} shows this effect as a function of the grid resolution. In the low resolution case the overestimation is about a factor of four. The interpolated mass converges to the total SPH mass within the analysed domain at $768^3$ resolution. Due to computational time constrains we choose to use the $512^3$ grid, which yields an error of less than 10 per cent in the mass. The overestimated mass generally results in brighter CO emission.

The probability density function of $^{12}$CO emission, in cases of normalised and non-normalised grid interpolation schemes, are compared in Fig.~\ref{appdxfig:Wcointerpol}. The normalised grid method yields higher peak emission value and a distribution of emission above 2 $\textrm{K}\,\textrm{km}\,\textrm{s}^{-1}$ that is shifted towards higher values. The overall luminosity is about 30 per cent higher in this case.

\section{Dependence on viewing angle} \label{appdx:viewangle}

\begin{figure}
\begin{center}
\includegraphics[trim=20 30 0 25, scale=0.48]{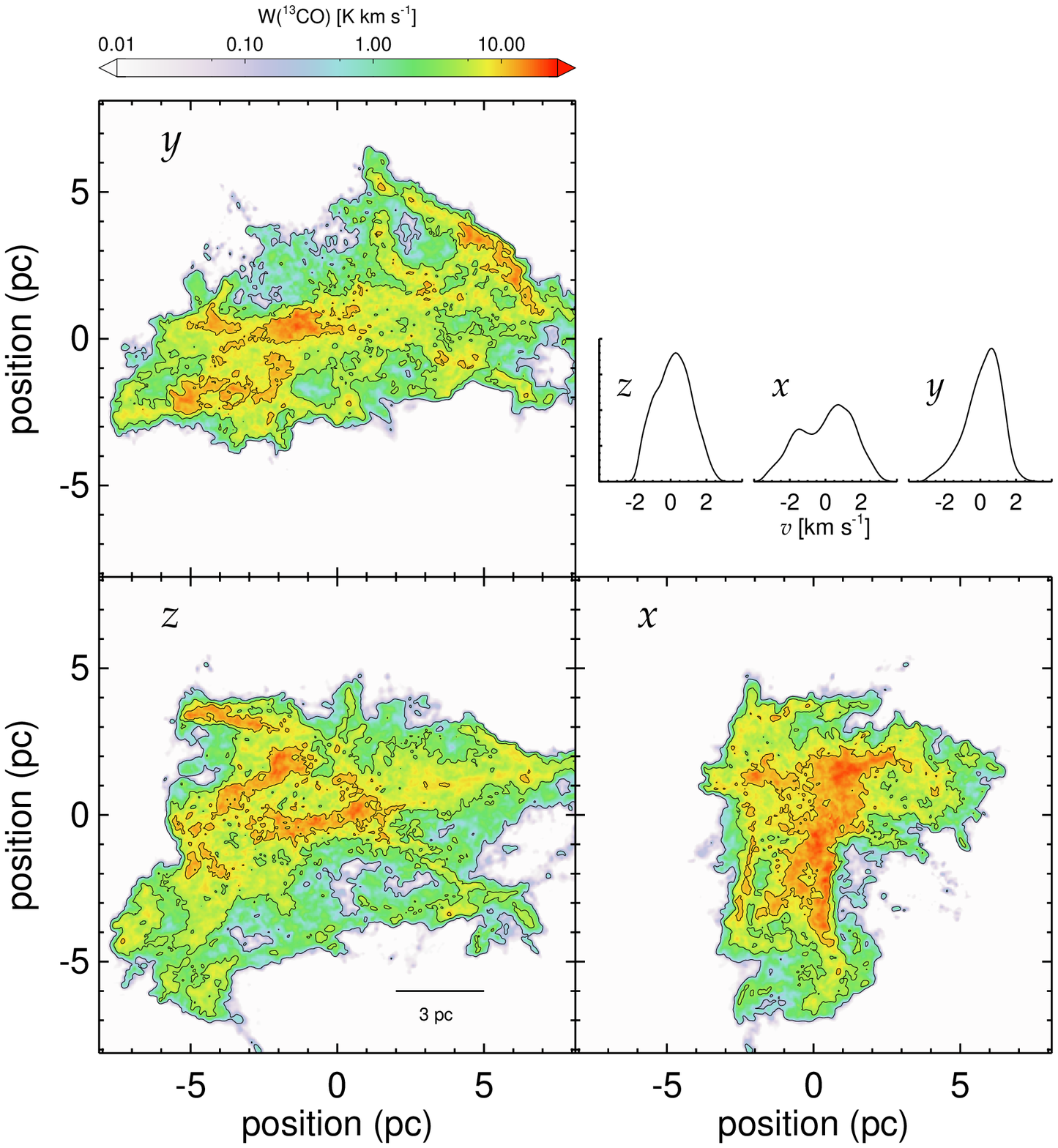}
\end{center}
\caption{Velocity channel integrated $^{13}$CO emission maps and the line profile at different viewing angles for the fiducial (d) cloud model. The mass estimated along the various sight lines are compared in Fig.~\ref{appdxfig:vamass}. The broad line along the $x$ direction clearly originates from diverging CO-bright regions, that are spatially overlapping in this projection. The contour lines show the 0.3 $\textrm{K km s}^{-1}$, 4 $\textrm{K km s}^{-1}$ and 10 $\textrm{K km s}^{-1}$ levels.}
\label{appdxfig:viewangle}
\end{figure}

\begin{figure}
\begin{center}
\includegraphics[trim=0 35 0 0, scale=0.5]{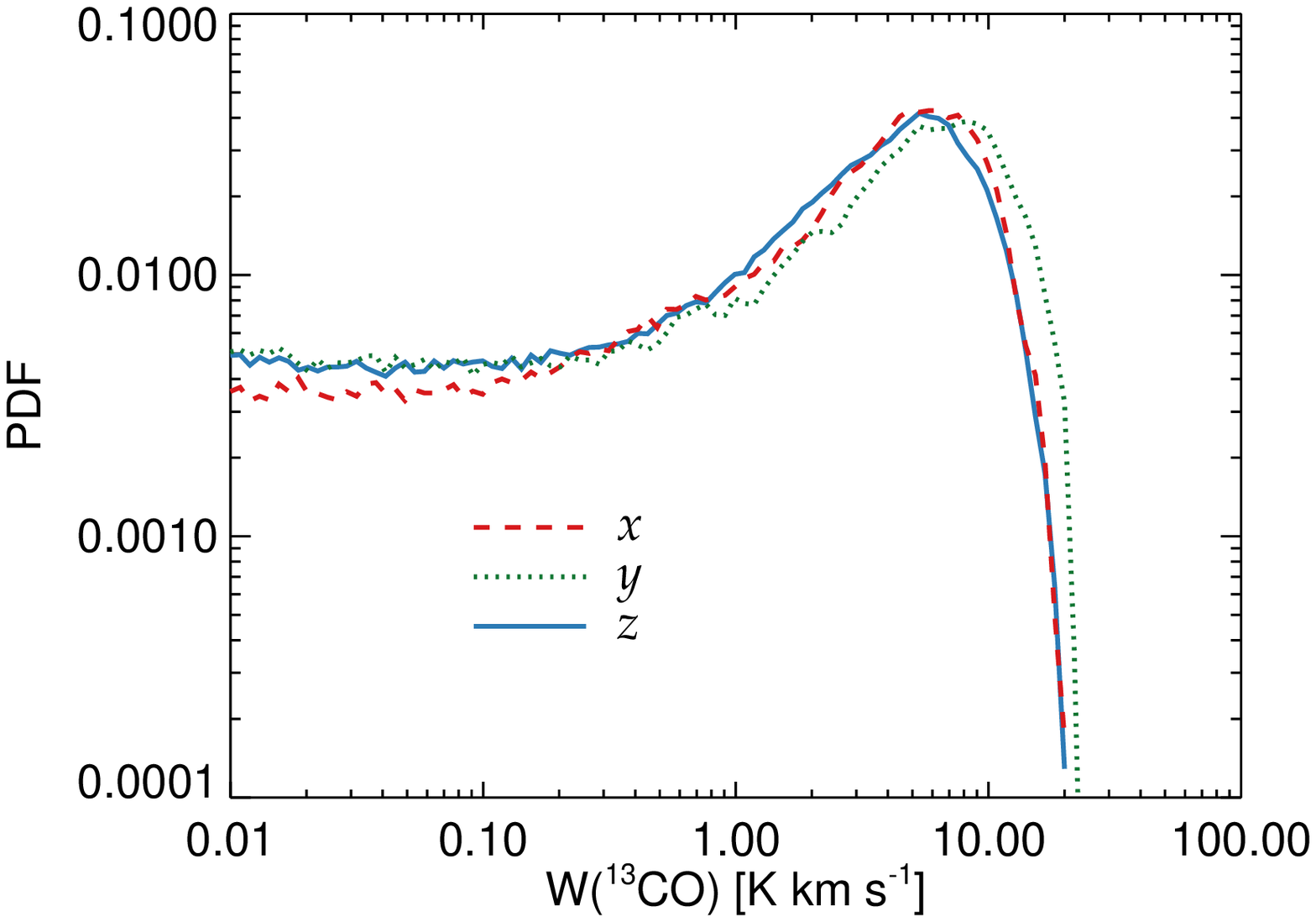}
\end{center}
\caption{Comparison of the $^{13}$CO integrated emission map PDFs at different viewing angle. The plotted curves correspond to the maps shown in Fig.~\ref{appdxfig:viewangle}.}
\label{appdxfig:WcointerpolWA}
\end{figure}

\begin{figure}
\begin{center}
\includegraphics[trim=90 176 0 20, scale=0.64]{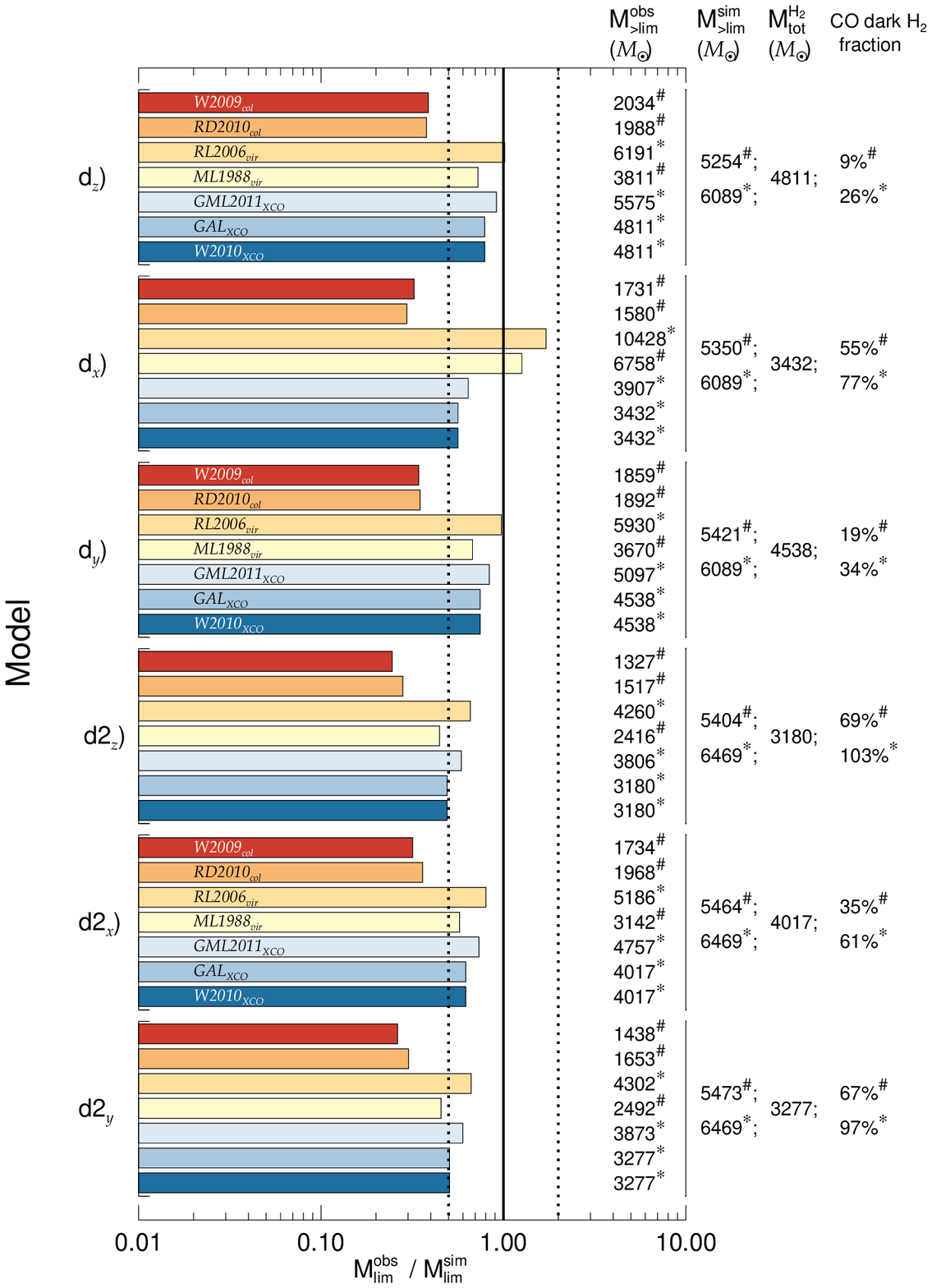}
\end{center}
\caption{Viewing angle dependence of the mass estimates for the fiducial cloud model (d) and for an independent realisation of the Milky Way-like cloud conditions (d2). For a given cloud, the viewing angle has only a minor effect on the mass estimates, while the cloud-to-cloud scatter might be larger. The exception is model d), viewed along $x$, where the double-peaked line profile affects both the virial and the $X_{\textrm{CO}}$ masses.}
\label{appdxfig:vamass}
\end{figure}

The analysis presented in the main body of the paper uses synthetic emission maps calculated along the $z$ axis of the simulations. Due to the initially isotropic gas density and velocity dispersion, we expect a weak dependence of cloud properties, e.g. the CO line emission, on the viewing angle. To test this expectation, we calculate the $^{12}$CO and $^{13}$CO ($J=1\rightarrow0$) emission along the $x$ and $y$ axes (see Fig.~\ref{appdxfig:viewangle} for the maps and Fig.~\ref{appdxfig:WcointerpolWA} for the integrated emission PDFs) of model d) and estimate the molecular mass using the observational methods discussed above (Fig.~\ref{appdxfig:vamass}).

The simulated cloud is elongated and shows very similar morphology and line shapes along the shorter axes ($z$ and $y$), but a more centrally concentrated emission and double peaked line profile along the long axis ($x$). The latter is due to overlapping, individual, CO-bright dense regions, which are moving away from each other along the $x$ axis.
The difference between the short axes and the long axis is reflected in the virial and $X_{\textrm{CO}}$ mass estimates, while the LTE column density estimate is not affected. Due to the larger width of the two component CO line, the virial mass is overestimated, while the $X_{\textrm{CO}}$ mass is underestimated.

We also test how the results change, when a different realisation (i.e. different shape, density distribution, etc.) of a Milky Way-like cloud (model d) is analysed. We set the parameters as in model d), but the turbulent velocity field is initialised with a different random seed. The $^{12}$CO and $^{13}$CO synthetic maps are calculated along the $x$, $y$, $z$ directions of the last snapshot before sink particle formation, and the molecular mass is estimated. The results are summarised in the lower half of Fig.~\ref{appdxfig:vamass}.
The methods give very similar results along the different sight lines. We find a larger cloud-to-cloud variation in the ability of methods to recover the molecular mass, although the difference is still within a factor of two.

\end{document}